\documentclass[journal]{IEEEtran}
\makeatletter
\long\def\@makecaption#1#2{\ifx\@captype\@IEEEtablestring%
\footnotesize\begin{center}{\normalfont\footnotesize #1}\\
{\normalfont\footnotesize\scshape #2}\end{center}%
\@IEEEtablecaptionsepspace
\else
\@IEEEfigurecaptionsepspace
\setbox\@tempboxa\hbox{\normalfont\footnotesize {#1.}~~ #2}%
\ifdim \wd\@tempboxa >\hsize%
\setbox\@tempboxa\hbox{\normalfont\footnotesize {#1.}~~ }%
\parbox[t]{\hsize}{\normalfont\footnotesize \noindent\unhbox\@tempboxa#2}%
\else
\hbox to\hsize{\normalfont\footnotesize\hfil\box\@tempboxa\hfil}\fi\fi}
\makeatother

\usepackage{cite}
\usepackage{amsfonts}
\usepackage{epsfig}
\usepackage{subfigure}
\usepackage{mathtools}          %loads amsmath as well
\usepackage{graphicx}
\usepackage{color}
\ifCLASSINFOpdf
   \usepackage{algorithm}
	\usepackage[noend]{algpseudocode}

\begin{document}

%\title{Ch-RRI: Improved Semi-Persistant Scheduling using Adaptive Selection Windows for C-V2X}
%\title{Differentiated localization for NR-V2X applications}
%\title{Decentralized Radio Resource Prediction and Scheduling for NR-V2X localization}
%\title{Minimizing AOI for improved NR-V2X Localization}
%Decentralized Age of Information Aware Radio Resource Management for Improved NR-V2X Cooperative Awareness
%\title{Optimizing Information Freshness for NR-V2X Cooperative Awareness}
%\title{Optimizing Information Freshness for Decentralized NR-V2X Networks}
\title{Adaptive RRI Selection Algorithms for Improved Cooperative Awareness in Decentralized NR-V2X}
%Authors List
%\thanks{This paper was presented in part at IFIP Networking 2021 \cite{IFIP_networking_paper}.}
\author{Avik Dayal$^1$\thanks{An earlier version of this paper was presented in part at the 2021 IFIP Networking Conference \cite{IFIP_networking_paper}. A preliminary version of this work appears as a chapter in Avik Dayal's Ph.D. Thesis \cite{dayal2021practical}.}\thanks{This research is supported by the Office of Naval Research (ONR) under MURI Grant N00014-19-1-2621.}, Vijay K. Shah$^2$, Harpreet S. Dhillon$^3$, and Jeffrey H. Reed$^3$ \\
$^1$ Johns Hopkins University Applied Physics Laboratory, Laurel, MD 20723, USA
\\
$^{2}$ Cybersecurity Engineering Department, George Mason University, VA, USA \\
$^3$Wireless@VT, Bradley Department of ECE, Virginia Tech, VA, USA \\

avik.dayal@jhuapl.edu, \{hdhillon, reedjh\}@vt.edu and vshah22@gmu.edu}
% The authors are with Bradley Department of Electrical and Computer Engineering, Virginia Tech, and affiliated with Wireless@VT Lab. (Email:\{biplavc, vijays, ad6db, reedjh\}@vt.edu)},

\date{June 2020}

\maketitle

%Main body starts

\begin{abstract}

Decentralized vehicle-to-everything (V2X) networks (i.e., C-V2X Mode-4 and  NR-V2X Mode-2) utilize sensing-based semi-persistent scheduling (SPS) where vehicles sense and reserve suitable radio resources for Basic Safety Message (BSM) transmissions at prespecified periodic intervals termed as \textit{Resource Reservation Interval} (RRI). 
Vehicles rely on these received periodic BSMs to localize nearby (transmitting) vehicles and infrastructure, referred to as \textit{cooperative awareness}. Cooperative awareness enables line of sight and non-line of sight localization, extending a vehicle’s sensing and perception range. In this work, we first show that under high vehicle density scenarios, existing SPS (with prespecified RRIs) suffer from poor cooperative awareness, quantified as \textit{tracking error}. Tracking error is defined as the difference between a vehicle’s true and estimated location as measured by its neighbors. To address the issues of static RRI SPS and improve cooperative awareness, we propose two novel \textit{RRI} selection algorithms -- namely, \textit{Channel-aware RRI} (Ch-RRI) selection and \textit{Age of Information (AoI)-aware RRI} (AoI-RRI) selection. Ch-RRI dynamically selects an RRI based on \textit{channel resource availability} depending upon the (sparse or dense) vehicle densities, whereas AoI-RRI utilizes a novel information freshness metric, called Age of Information (AoI) to select a suitable RRI. Both adaptive RRI algorithms use SPS for selecting transmission opportunities for timely BSM transmissions at the chosen RRI. System-level simulations demonstrate that both proposed schemes outperform the SPS with fixed RRI in terms of improved cooperative awareness. Furthermore, AoI-RRI SPS outperforms Ch-RRI SPS in high densities, whereas Ch-RRI SPS is slightly better than AoI-RRI SPS in low densities.

\end{abstract}

\begin {IEEEkeywords}
NR-V2X, NR-V2X Mode-2, Age-of-Information, Dynamic Spectrum Access, Radio Resource Management
\end{IEEEkeywords}
\section{Introduction}
Modern vehicles have been equipped with a plethora of sensors to assist with autonomous capabilities \cite{mmwave_IOV_IOTJ_ghafoor}. The drawback is that most sensors, including cameras, radars, and LIDAR, are limited to line of sight (LOS) visibility and are constrained in their scope \cite{Collective_perception_Thandavarayan_2019}. This LOS constraint is a major motivation for the development of Vehicle-to-Everything (V2X) communications. V2X communications enables both LOS and NLOS exchange of sensor data with neighboring vehicles, thereby increasing the range of a vehicle's \textit{awareness} of surrounding vehicles and infrastructure \cite{V2V_safety_coop_perception}. %To this end, the Third Generation Partnership Project (3GPP) has developed radio access technologies (RATs) based on Long Term Evolution (LTE), dubbed C-V2X, to facilitate V2X communications. Since the presence of cellular infrastructure cannot be assumed, 3GPP has proposed using sidelink communications for the dissemination of each vehicles state information. 

The Third Generation Partnership Project (3GPP) Release 16 has introduced a V2X communications technology that uses the 5G New Radio (NR) air interface, referred to as 5G NR-V2X. NR-V2X includes enhancements to the 3GPP's previous cellular V2X communications technology (C-V2X) based on the Long-Term Evolution (LTE) standard. 
NR-V2X offers two modes of operation, termed Mode-1 and Mode-2, that use sidelink communications, which is direct communication between users or vehicles without data passing through the gNodeB \cite{Naik_survey_5G}. %NR-V2X offers two modes of operation termed Mode-1 and Mode-2 that allow vehicles and users to communicate directly with partial or no gNodeB coverage. NR-V2X offers two modes of operation termed Mode-1 and Mode-2 that allow vehicles and users to communicate directly with partial or no gNodeB coverage, called sidelink communications. Mode-1 NR-V2X employs a centralized scheduling approach, where a gNodeB schedules links for two or more vehicles to communicate directly. 
NR-V2X Mode-1 employs a centralized scheduling approach, where a gNodeB schedules and assigns sidelink radio resources for two or more vehicles to communicate and exchange data directly. NR-V2X Mode-2 assumes communications to occur outside the coverage of an gNodeB; therefore, every vehicle uses a sensing-based semi-persistent scheduling (SPS) protocol to sense and reserve sidelink radio resources. SPS was introduced as part of the Release 14 C-V2X Mode-4 standard, the predecessor to NR-V2X Mode-2 \cite{molina2020comparison}. Since cellular connectivity can not be assumed ubiquitous, NR-V2X Mode-2 is considered as the baseline mode for NR-V2X.

\par In NR-V2X Mode-2 SPS, vehicles sense and select suitable (available) radio resources to transmit packets at predefined fixed time intervals, termed, Resource Reservation Interval (RRI).
%\par In the NR-V2X Mode-2 standard, resources are reserved through sensing-based SPS, where vehicles sense radio resources (or channel medium) and select suitable (underutilized) radio resources for the transmission of packets at prespecified fixed time intervals, termed, \textit{Resource Reservation Interval} (RRI). 
Equivalently, RRI is defined as the inter-transmission time interval between two consecutive transmissions. In the Release 14 C-V2X standard, RRIs equal to or below $100$ ms were restricted to $20$ ms, $50$ ms, or $100$ ms \cite{magazine_paper}. Release 16 NR-V2X Mode-2 provides more flexibility by allowing any integer RRI between $1$ and $99$ ms for any RRIs below $100$ ms \cite{Tutorial_5G_NR_V2X}. 
\par The primary use case for NR-V2X Mode-2 SPS is the dissemination of basic safety messages (BSMs), which carry time-sensitive state information such as a transmitting vehicle's speed, heading, and location \cite{ETSI_EN_302}. Receiving vehicles can use these BSMs for \textit{cooperative awareness}, including the localization and trajectory prediction of neighboring vehicles and infrastructure. Cooperative awareness can enable safety-critical applications such as forward collision warnings \cite{milanes2013cooperative}, blind spot/lane change warnings \cite{Huang_IEEENetwork}, and is likely one of the key requirements for autonomous driving \cite{rasouli2019autonomous}. %This is mainly because BSMs facilitate \textit{accurate positioning} or \textit{localization} of neighboring vehicles. 
One metric that can quantify cooperative awareness performance is tracking error, the difference between a vehicle's actual and estimate location (via most recent BSM) by its neighboring vehicles. Tracking error, as opposed to conventional communication metrics such as packet delivery ratio (PDR) and throughput, captures the impact of multiple lost or outdated BSMs on a vehicle's awareness and is thus a more appropriate performance measure for this application.%Tracking error is preferred to more conventional communication metrics, such as packet delivery ratio and throughput, that do not capture the impact of lost or outdated BSMs on a vehicle's awareness.

\par Since BSMs carry time-sensitive information, outdated BSMs (due to large RRIs) and/or lost BSMs (due to channel congestion) negatively impact the performance of cooperative awareness applications, mainly due to erroneous localization of neighboring vehicles. Therefore, it is critical that the state information in each vehicles transmission be \textit{fresh}, as stale information can compromise the aggregate awareness of the vehicular network. The \textit{freshness} of information at receiving vehicles can be measured using Age-of-Information (AoI). AoI is the time elapsed since new state information has been generated and is a promising metric to measure the freshness of system state information. Results in \cite{Biplav_VTC_2020} have found that AoI has a strong correlation with tracking error in vehicular networks, which motivates further study of AoI.  %Since the NR-V2X Mode-2  incorporates sensing measurements as part of the standard and allows reservations for transmissions to be made in time and frequency, the sensing history can be used to help vehicles choose more optimal transmission opportunities, yielding lower AoI and a better safety experience. 

\par In this work, we explore the deficiencies of the NR-V2X Mode-2 SPS using static RRIs in terms of tracking error and propose algorithms that choose adaptive RRIs to improve the overall cooperative awareness performance of NR-V2X Mode-2. We show in this paper that SPS with static RRIs (e.g. $50$ and $100$ ms) has cooperative awareness limitations in NR-V2X networks. 
Consider these scenarios: (i) \textit{high vehicle density scenarios} -- the NR-V2X network performance suffers from congestion with a large number of lost packets, leading to an increased tracking error, and (ii) \textit{low vehicle density} -- the spectrum resources would be under-utilized in time. The tracking error performance in low density scenarios can improve with a smaller RRI, which ensures frequent location updates without fear of channel congestion. We introduce two adaptive RRI algorithms, termed Channel-aware RRI (Ch-RRI) selection and AoI-aware RRI (AoI-RRI) selection. Ch-RRI uses channel occupancy measurements to find the smallest RRI possible without increasing congestion. Motivated by the benefits of reducing AoI, we propose AoI-RRI, an AoI-aware RRI selection algorithm for NR-V2X Mode-2  that uses the average AoI observed by neighboring vehicles to select an optimal RRI. 

%Consider the following instances: (i) \textit{high vehicle density scenarios} -- C-V2X networks would likely suffer from overly congested radio channels (and thus, large number of lost or dropped BSM packets), which will lead to increased tracking error and thus, degraded cooperative awareness, and (ii) \textit{low vehicle density} -- The radio resource are under-utilized in such low density scenarios. The cooperative awareness can be greatly improved by choosing lower value of RRIs (e.g., $20$ ms), as lower value of RRIs will improve the timeliness of BSMs without compromising the channel congestion (as the channels can support lower RRIs for fewer vehicles). Furthermore, notice that due to vehicle mobility and other contextual factors, the vehicle densities and available channel resources may change over time even for a given C-V2X scenario. 
% Thus, \textit{there is a need for an optimal scheduling algorithm that determines suitable value of RRI at each vehicle at any given time slot, such that the BSM timeliness is ensured without compromising the channel congestion under any considered C-V2X network scenario.} 

%The successor to Mode-4 C-V2X in the 5G new radio (NR) standard, named Mode-2(a), is expected to be based on the same sensing based SPS in Release 14, so we anticipate that our Ch-RRI algorithm will also be applicable to the new NR-V2X standard .

In this work, we make following key contributions:

\begin{itemize}
%\item We demonstrate that NR-V2X Mode-2 insufficiently distributes radio channel resources, negatively impacting the timely tracking error and age of information of information near neighboring vehicles. To our knowledge, this is the first study of age for the NR-V2X standard. 
\item We show that NR-V2X Mode-2 SPS with static RRIs suffers severely from under- and over-provisioning of radio resources (depending upon the vehicle densities). This in turn negatively impacts the timely successful delivery of BSMs and compromises the cooperative awareness of NR-V2X Mode-2 SPS.  

\item To address the limitations of SPS with static RRIs, we propose Ch-RRI SPS, which is SPS powered by a \textit{channel aware} RRI selection algorithm. The Ch-RRI algorithm uses channel occupancy measurements and chooses the minimum possible RRI spacing between transmissions without increasing packet loss due to congestion.  Ch-RRI SPS uses this RRI spacing to select radio resources for BSM transmissions.

\item In a similar vein to Ch-RRI SPS, we develop AoI-RRI SPS, SPS powered by a \textit{AoI aware} RRI selection algorithm. The AoI-RRI algorithm uses neighborhood age and channel resource measurements to choose an age optimal RRI iteratively. %This allows vehicles to intelligently reserve resources for transmissions. 
%\item We propose two adaptive RRI protocols, termed Ch-RRI and AoI-RRI that ensure BSM timeliness and thus, improves the cooperative awareness performance of NR-V2X Mode-2. Ch-RRI allows each vehicle to dynamically adjust its RRI at each time slot, and judiciously utilize radio resources for BSM transmissions while accounting for varying C-V2X vehicle traffic scenarios. AoI aware NR-V2X uses measurements of the neighborhood age and channel resources to choose a age optimal RRI when selecting resources. This allows vehicles to intelligently allocate resources for transmissions across various highway densities. 

%\item Experiments developed on our NR-V2X simulator show that the proposed Ch-RRI SPS and AoI-RRI SPS outperform conventional SPS in terms of tracking error, across varying vehicle traffic scenarios. 
\item Experiments developed on our NR-V2X simulator show that when compared to NR-V2X Mode-2 SPS with $20$ ms, $50$ ms, and $100$ ms RRI, the proposed Ch-RRI and AoI-RRI SPS demonstrate a substantial reduction of tracking error. Ch-RRI as compared to AoI-RRI shows a lower AoI and tracking error in low density highway scenarios, and is faster in converging to a local minimum AoI. However, the average AoI and tracking error of AoI-RRI is lower at higher densities.%by $51.27$\%, $51.20$\%, and $75.41\%$ for $80$ vehicles/km. 
%The results are consistent for other considered C-V2X scenarios. Furthermore, Ch-RRI performs well in terms of packet delivery ratio in all considered scenarios.

\end{itemize}
%An earlier version of this paper was presented in part at the 2021 IFIP Networking Conference \cite{IFIP_networking_paper}. A preliminary version of this work appears as a chapter in Avik Dayal's Ph.D. Thesis \cite{dayal2021practical}.
The organization of this paper is as follows:  Section \ref{Background} discusses related works. Section \ref{Section: Mode 2 NR-V2X Sidelink} discusses cooperative awareness, the tracking error metric, and the NR-V2X Mode-2 standard. In Section \ref{Section: Ch-RRI} we illustrate the limitations of traditional fixed RRI SPS and present the algorithmic details of Ch-RRI SPS. Section \ref{Section: AoI aware SPS} provides a brief description of AoI, and discusses the details of AoI-RRI SPS. Finally, Section \ref{Section- Simulation Description and Results} discusses the results and provides an outlook for future work, followed by concluding remarks in Section \ref{Section-Conclusion}.
\section{State-of-the-art}
\label{Background}
%\par Although there has been a lack of work focused on using AoI in NR-V2X Mode-2  and Mode-4 C-V2X (the precursor to the NR-V2X standard), 
There has been a surge of research in the area of simulating and enhancing SPS for C-V2X Mode-4 \cite{magazine_paper, IOTJ_2016_V2X_Chen,Nabil_SPS,Bazzi_CV2X, IOTJ_SPS, bazzi2019congestion, Peng_TWC_AOI_optimized_mac, Bexmenov_SPS_Aperiodic, Analytical_CV2X} and more recently, for NR-V2X Mode-2 SPS \cite{campolo20195g,Zoraze_NRV2X_Mode2_overview, Bartoletti_Access_2021,Todisco_Access_2021}. The research performed in \cite{IOTJ_2016_V2X_Chen} simulated and provided a baseline for semi-persistent scheduling across various densities for highway and urban environments. Research in \cite{Nabil_SPS}, \cite{Bazzi_CV2X}, and \cite{molina2020comparison} varied the probability of reselection, selection window, and RRI in SPS and measured the impact on the packet delivery ratio (PDR). The authors of \cite{IOTJ_SPS} concluded that using short term sensing before resource selection reduced packet collisions and improved SPS performance. The work in \cite{bazzi2019congestion} and \cite{Lee_TENCON_2020} was amongst the first to adjust the transmission power and RRI accordingly, to improve the overall performance of SPS. To address the scheduling overhead of SPS, \cite{Bexmenov_SPS_Aperiodic} has proposed using single shot transmissions as part of SPS for safety critical messages that have an immediate scheduling need. Although much of the research for semi-persistent scheduling is simulation driven, \cite{Analytical_CV2X} has provided an analytical model for semi-persistent scheduling in vehicular networks. Recently, the work in \cite{campolo20195g,Zoraze_NRV2X_Mode2_overview, Bartoletti_Access_2021,Todisco_Access_2021} has simulated the relatively new NR-V2X Mode-2 SPS, specifically looking at the impact of the flexible numerology on SPS. A potential drawback of these works is the focus on improving the coverage (typically measured through PDR) of C-V2X Mode-4 and NR-V2X Mode-2, and ignoring the cooperative awareness performance of the vehicular network.
\par AoI has recently emerged as a popular metric for quantifying the performance of scheduling and decentralized radio resource management in vehicular networks. In particular, there has been a lot of research done in optimizing the broadcast rate to reduce the AoI in networks that use Dedicated Short Range Communications (DSRC), a competing V2X technology based on the 802.11p standard. \cite{kaul2012real,baiocchi2020age,vinel2015vehicle,AOI_DCC,yates2018age,Kosta_2019_AOI_packet_management,Maatouk_2020_AOI}. Although there are notable differences between the two standards, changing the broadcast rate can be seen as the analogous equivalent to using the RRI to minimize the AoI, and merits discussion. The authors in \cite{kaul2012real} developed a decentralized algorithm that iterated the broadcast rate based on AoI measurements for a DSRC network. The work in \cite{baiocchi2020age} proposed a model that uses a vehicular network's connectivity graph to show the relationship between the average system AoI and vehicle density and broadcast intervals. AoI was used to evaluate platooning applications for convoys of vehicles in \cite{vinel2015vehicle} and \cite{AOI_DCC}. Furthermore, \cite{Kosta_2019_AOI_packet_management} and \cite{Maatouk_2020_AOI} investigated the effect of the backoff window size in CSMA on the AoI.
\par Although the SPS protocol in C-V2X Mode-4 and NR-V2X Mode-2 is relatively new, there has been some recent work in applying AoI to SPS. The authors in \cite{Peng_TWC_AOI_optimized_mac} analyzed the AoI of Release 14 C-V2X Mode-4 and proposed a piggyback collaboration method to help decrease half duplex errors, thereby reducing the AoI. More recently, \cite{Cao_WCNC_AOI_2022} looked at the effect of RRI spacing on AoI performance in the NR-V2X Mode-2 standard, and found optimal RRI values across vehicle densities. Note that although the Release 14 C-V2X Mode-4 SPS is similar to Release 16 NR-V2X Mode-2 SPS, there are a few key differences between the two standards. Two differences that are particularly relevant to this work are the change in ranking of slot resources (discussed in Section \ref{Section: Ch-RRI}) and the inclusion of any integer RRI between 1 and 99 ms\footnote{Please refer to \cite{Tutorial_5G_NR_V2X} for a detailed discussion on the differences between C-V2X Mode-4 and NR-V2X Mode-2.}. These changes can enable higher transmission rates while preventing collisions from choosing the same slot resource \cite{Zoraze_NRV2X_Mode2_overview}. While there have been recent works evaluating the overall performance of Release 16 NR-V2X SPS, to the best of our knowledge, this is the first work that evaluates the AoI and cooperative awareness performance for NR-V2X SPS and the first work to directly use AoI measurements to optimize the RRI for SPS.

\section{Cooperative Awareness and NR-V2X Mode-2}
\label{Section: Mode 2 NR-V2X Sidelink}
% \begin{figure*}[h]
% \centering
% \includegraphics[width=6.25 in]{images/NR-V2X_SPS_sensing_v2.png}
% \caption{Illustration of Collision Risky scenario caused by a large tracking error.}
% \label{mode_2_sensing_fig}
% \centering
% \end{figure*} 

% \begin{figure*}[!h]
% % \vspace{-0.05in}

% %Illustration of the sensing window, selection window, resource reservation interval, and scheduling for SPS
% %\subfigure[\label{SPS_illustration}]{
% %\epsfig{figure=images/NR-V2X_SPS_sensing_v2.png,width=4.25 in,  keepaspectratio}}
% \centering
% \subfigure[\label{Mode_2a_algorithm}]{
% \epsfig{figure=images/NR_V2X_Mode2a_v3.png,width=2.9 in,  keepaspectratio}}
% \subfigure[\label{adaptive_SPS_algorithm_fig}]{
% \epsfig{figure=images/adaptive_SPS_algorithm_figv6.png, width=2.9 in,  keepaspectratio}}
% %\subfigure[\label{adaptive_SPS_algorithm}]{

% \caption{(a) Flowchart of NR-V2X Mode-2 and (b) Flowchart of Ch-RRI.}
% %\caption{(a) Illustration of the sensing window, selection window, RRI, and scheduling for SPS and (b) Illustration of Collision Risky scenario caused by a large tracking error.}\vspace{-0.2in}
% \end{figure*}

This section discusses cooperative awareness, the tracking error metric, and the NR-V2X Mode-2 standard for sidelink communications from the physical layer structure to the semi-persistent scheduling algorithm that vehicles use to find and reserve suitable transmission opportunities. %3GPP Release 16 NR-V2X is based on 3GPP Release 12 device-to-device communications, first introduced for proximity services (ProSe). 3GPP Release 14 was the first standard to use D2D communications for V2X communications(termed LTE-V2X).

\subsection{Cooperative Awareness and Tracking Error}
\label{Subsection:Tracking Error}
%The scenario assumes that vehicle $v$ receives messages transmitted by vehicle $u$ and uses these messages to $localize$ vehicle $u$'s position. 
Since cooperative awareness depends on the sharing of time-sensitive information with other vehicles via BSMs, conventional communication metrics such as latency, throughput, and packet delivery ratio (PDR) greatly impact a vehicle's awareness. However, these metrics alone cannot capture the impact of multiple lost or outdated BSMs on a vehicle's awareness, as illustrated by the following example. Consider a fast moving vehicle travelling at $140$ km/hr ($38.89$ m/s) while transmitting with a high RRI ($100$ ms or more). Neighboring vehicles receive location packets reliably, but because there is at least $100$ ms between packets, the localization or \textit{tracking} error would be at least $3.89$ m. Likewise, a low RRI ($50$ ms or less) would yield a tracking error of at least $1.94$ m, though this tracking error could increase if packets are dropped due to channel congestion. Thus, as the above example shows, a tracking error metric can provide a better context to the cooperative awareness performance in vehicular networks.
\par
\begin{figure}[h]
\centering
\includegraphics[width=\columnwidth]{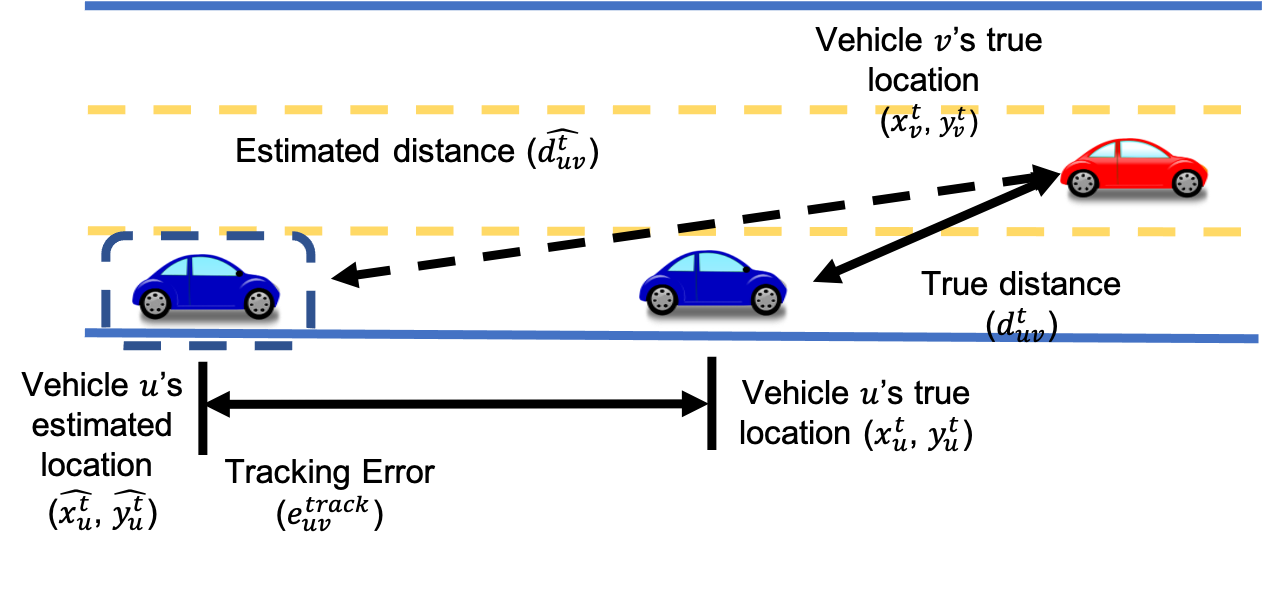}
\caption{Illustration of latency induced tracking error in a vehicular network.}
\label{tracking_error_fig}
\centering
\end{figure}

 Illustrated in Fig. \ref{tracking_error_fig}, tracking error ($e^{track}_{uv}$) in the context of NR-V2X is defined as the difference in the transmitting vehicle's (vehicle $u$) true location, and $u$'s estimated location by neighboring receiving vehicle $v$. We assume that the receiving vehicle $v$ is transmission range of $u$ and time $t^{'}$ is the generation time of the transmission from $u$ to $v$. The tracking error is calculated as:

%when BSM is received, the position of $u$ has changed to $x^t_u$. The tracking error at $v$ while tracking $u$ at time $t$ is given by:

% \begin{equation}\label{TE_eq}
%     e^{track}_{uv} = |x^t_u - x^{t'}_u|
% \end{equation}

\begin{equation}
   e^{track}_{uv} = \sqrt{ (x_{u}^{t} - \hat{x}_{uv})^2 + (y_{u}^{t} - \hat{y}_{uv})^2 },
    \label{TE_eq}
\end{equation}
where ($x^t_u$, $y^t_u$) is $u$'s true current 2-D location at time $t$ and ($\hat{x}_{uv}$, $\hat{y}_{uv}$) is $u$'s previous location from time $t^{'}$ at vehicle $v$ from the last transmission received from $u$. %We consider $x$-coordinate to calculate the tracking error as the vehicle moves in x-direction only\footnote{For the ease of presentation, we consider a simple tracking error (TE) model with no lateral movements across lane.}.

% Note vehicle $u$ may not see the same TE in tracking vehicle $v$, because $v$'s BSM may have been generated at different times, say $t'' \neq t'$. 
%and also, $v's$ location, speed etc. may be different to that of $u$ (thus, incurring significantly different $x^{t''}_v$).
The tracking error has a large impact on the cooperative awareness in vehicular networks because 

\begin{enumerate}
    \item Every transmission has a non-zero propagation delay between transmission and successful reception and the transmitting vehicle would also likely have moved during the transmission and reception times. 
    \item The RRI (or inter-packet transmission interval) can cause a large tracking error, especially at higher speeds. %For example, in the scenario depicted in Fig. \ref{tracking_error_fig}, a $100$ ms RRI yields at least a $100$ ms gap between two transmissions from transmitting vehicle $u$ and receiving vehicle $v$. Assuming the transmitting vehicle travels at 70 km/hr, a $100$ ms RRI creates a 1.9 m tracking error. 
    
    %\item Consider a certain vehicle $u$ broadcasts its BSM at certain periodic time intervals or RRIs, say $100$ ms. It means there is at least $100$ ms inter-reception delay between two consecutive BSMs from that vehicle $u$ received at a neighboring vehicle $v$. Again during this time interval, the vehicle $u$ would like move a certain distance, incurring significant TE.
    \item Channel congestion could lead to several lost and delayed packets, further deteriorating tracking error. In the example of a vehicle traveling at $140$ km/hr with a $50$ ms RRI in a congested environment, two consecutive missed packets could cause a $5.833$ m tracking error.
\end{enumerate}
A lower tracking error at receiving vehicle $v$ implies that  $v$ can accurately position (or \textit{localize}) $u$. Note from the above discussion that a lower RRI can decrease the tracking error but can also increase the number of lost packets due to channel congestion. 

\subsection{Physical Layer Structure for the NR-V2X Sidelink}
Rel. 16 NR-V2X sidelink is designed to operate in two different frequency ranges, from $0.410$- $7.125$ GHz (FR1) and $24.25$- $52.5$ GHz (FR2) \cite{Tutorial_5G_NR_V2X}. Though both frequency ranges are supported in Rel. 16, NR-V2X is expected to operate in the FR1. The NR-V2X sidelink uses cylic prefix orthogonal frequency modulation (CP-OFDM) and supports multiple subcarrier spacings of $15$, $30$, and $60$ kHz. 
\subsubsection{Resource Structure in Time Domain}
In the NR-V2X sidelink, resources in the time domain are made up of frames, subframes, and slots. Every frame is made up of $10$ subframes, and each frame time length is typically $10$ ms \cite{Naik_survey_5G}. Subframes are typically $1$ ms in length, and broken down into slots. A slot consists of $4$ OFDM symbols, which means that the length of each slot depends on the chosen subcarrier spacing. A subcarrier spacing of $15$ kHz corresponds to a slot length of $1$ ms, while a $60$ kHz subcarrier spacing corresponds to a $0.25$ ms length \cite{Challenges_CV2X_Gyawali_2021}. For the purposes of the work, we assume NR-V2X Mode-2 SPS is inter-operable with C-V2X Mode-4, and take assume a subcarrier spacing of $15$ KHz
\subsubsection{Resource Structure in Frequency Domain}
The smallest schedulable unit of frequency are resource blocks (RB). In the 5G NR standard, a RB is made up of $12$ equally spaced subcarriers. The bandwidth of each RB depends on the subcarrier spacing value. RBs are combined sequentially to form subchannels. In NR-V2X, the data packets, or Transport Blocks, are transmitted on one or more subchannels. In each subframe, alongside every TB, is a sidelink control information block with modulation and coding scheme used 
\cite{molina2020comparison}.   
\subsection{NR-V2X Mode-2 Semi-persistent Scheduling}
\begin{figure}[h]
\centering
\includegraphics[width=.75 \columnwidth]{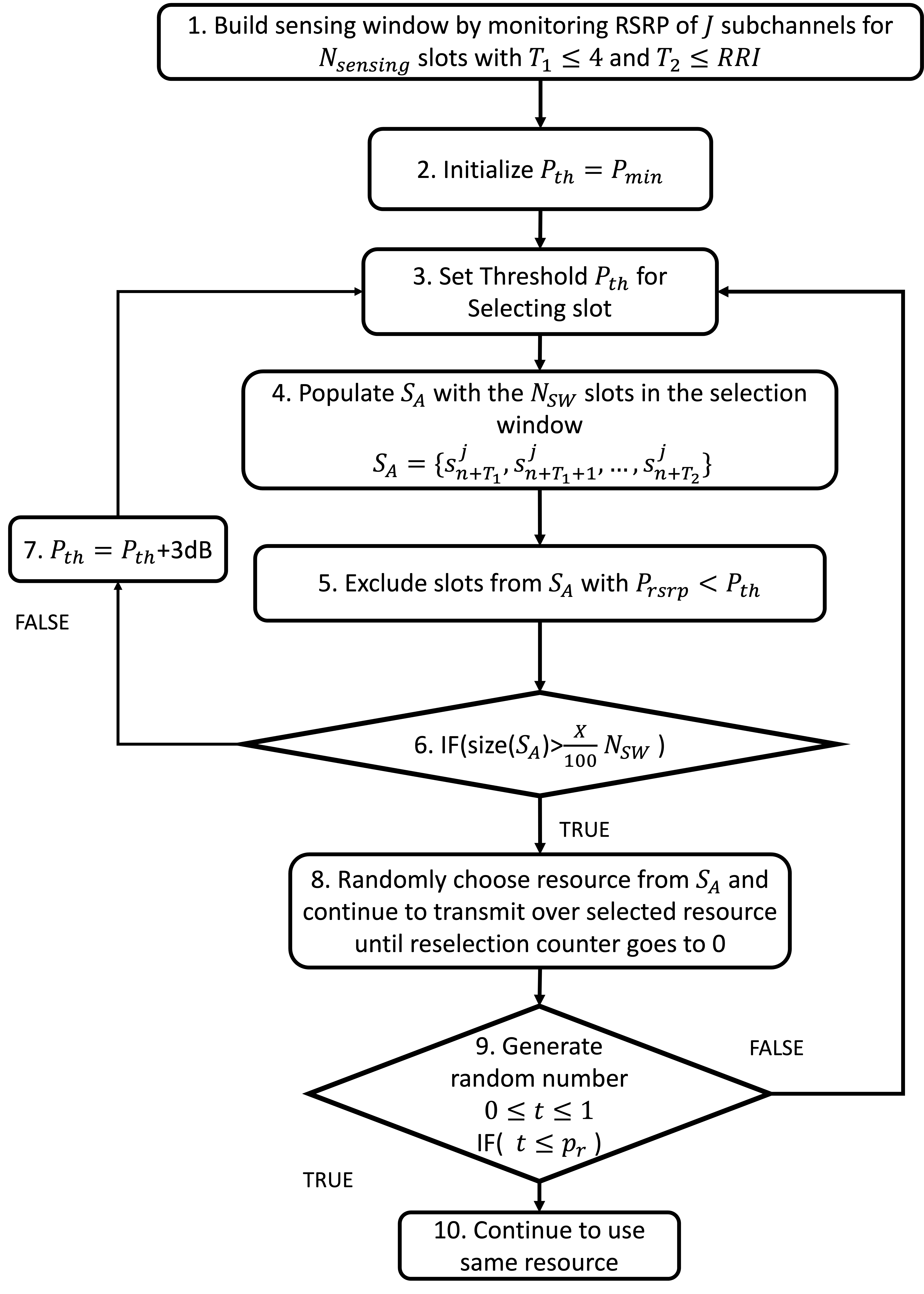}
\caption{Flowchart of SPS Algorithm.}
\label{SPS_algorithm_fig}
\centering
\end{figure}
At the MAC layer, the NR-V2X sidelink utilizes semi-persistent scheduling (SPS) that uses sensing to determine suitable semi-persistent transmission opportunities, i.e., set of slots, for BSM transmission. Fig. \ref{SPS_algorithm_fig} depicts the SPS algorithm\footnote{Please refer to \cite{magazine_paper} and \cite{Tutorial_5G_NR_V2X} for a detailed discussion on NR-V2X and SPS.}, and is explained below. We use $s_{i}^{j}$ to refer to a single-slot resource where $i$ is the slot index and $j$ is the subchannel index of $J$ total subchannels. %See Fig. \ref{SPS_illustration} for the illustration of slot and other key terminologies (e.g., sensing window) related to SPS.

\begin{itemize}
    
    \item \textbf{Sensing (Step 1):} Each vehicle continuously monitors the slots by measuring the reference signal received power (RSRP) across all $J$ subchannels; and stores sensing measurements for a prespecified last $N_{\rm sensing}$ slots, known as the \textit{sensing window}. Let slot $s_n$ denote the first slot after the sensing window. Then we can write the sensing window at $s_n$ as the following set of single-slot resources for the $j^{th}$ subchannel: $\left[s_{n-N_{\rm sensing}}^{j}, \dots, s_{n-1}^{j}\right]$.
    
    %\Vijay{Not clear -- Thus, if the subframe index at $t=0$, i.e. after measuring $N_{\rm sensing}$ subframes but before reservation and transmission, is $n$, the sensing window consists of the following set of single-subframe resources for the $j^{th}$ subchannel: $\left[sf_{n-N_{\rm sensing}}^{j}, \dots, sf_{n-1}^{j}\right]$.}
    
    \item \textbf{Identifying Available Resources (Steps 1-8):} 
    Each vehicle initializes a \textit{selection window} with a set of consecutive candidate slots (See Step $1$). $T_1 \leq 4$ and $T_2 \leq $ \textit{RRI} are the start and end slots for the selection window. %\Vijay{Let $N_{SW} = (T_2 - T_1)$ denotes the number of slots in the selection window.} 
    The \textbf{RRI} refers to the time interval between two consecutive BSM transmissions\footnote{In NR-V2X Mode-2 SPS, each vehicle selects an RRI at the time of resource selection, although the standard leaves the selection of RRI up to the user\cite{Tutorial_5G_NR_V2X}.}. %(See Fig. \ref{SPS_illustration} for illustration). 
    Each vehicle utilizes $N_{\rm sensing}$ slots (obtained in Step $1$) for identifying and subsequently, selecting the available slot  within the selection window for BSM transmission
    as follows. 
    \begin{enumerate}
        \item The vehicle sets -- (i) the RSRP threshold, $P_{th}$, to a minimum RSRP value, $P_{min}$ (Steps $2$-$3$) and (ii) initializes set $S_A$ as all slots in the selection window, i.e., $S_A=[s_{n+T_1}, s_{n+T_1+1}, \cdots, s_{n+T_2}]$ (See Step $4$). 
         \item As shown in Step $5$ of Fig. \ref{SPS_algorithm_fig}, the vehicle excludes all candidate subframes from set $S_A$ if one of the following conditions are met -- (i) the vehicle has not monitored the corresponding candidate subframe in the sensing window (i.e., $N_{\rm sensing}$) due to the half duplex exclusion criteria and (ii) the RSRP measurement for corresponding candidate subframe is higher than $P_{th}$. The RSRP exclusion criteria for the $i^{th}$ subframe (for $j$th sub-channel) in the selection window can be written as 
        
        %$RSRP(\sum_{k=0}^{}sf_{n+T_1+i-N_{\rm sensing}+k \cdot RRI}^{j})$
        \begin{equation} \label{RSRP_calculation1}
            \hspace*{-1.5cm}     RSRP\left(s_{n+T_1+i-N_{\rm sensing}}^{j}\right)\geq P_{th}
        \end{equation}
        %where $\mathcal{K} = \frac{N_{\rm sensing}}{RRI}$. 
        %Depending upon the value of $RRI$, value of $\mathcal{K}$ changes. 
        %If $RRI=100$ ms and $N_{\rm sensing} = 20$ ms and $i = 4$, then, we look at RSRP value across following $10$ subframes -- $\{4, 24, 44, \dots, 94\}$ and divide it by $\mathcal{K} = \frac{100}{20} = 5$ to find the average RSRP.
        % \item As shown in Step $5$ of Fig. \ref{SPS_algorithm_fig}, the vehicle excludes all candidate slots from set $S_A$ if one of the following conditions are met -- (i) the vehicle has not monitored the corresponding candidate slot in the sensing window (i.e., $N_{\rm sensing}$) and (ii) the linear average RSRP measurement for corresponding candidate slot is higher than $P_{th}$. The RSRP exclusion criteria for the $i^{th}$ slot (for $j$th sub-channel) in the selection window can be written as 
        
        % %$RSRP(\sum_{k=0}^{}sf_{n+T_1+i-N_{\rm sensing}+k \cdot RRI}^{j})$
        % \begin{equation} \label{RSRP_calculation1}
        %     \hspace*{-1.5cm}     \frac{1}{\mathcal{K}}\sum_{k=0}^{\mathcal{K}}RSRP\left(s_{n+T_1+i-N_{\rm sensing}+k \cdot RRI}^{j}\right)\geq P_{th}
        % \end{equation}
        % where $\mathcal{K} = \frac{N_{\rm sensing}}{RRI}$. 
        % %Depending upon the value of $RRI$, value of $\mathcal{K}$ changes. 
        % If $RRI=100$ ms and $N_{\rm sensing} = 20$ ms and $i = 4$, then, we look at RSRP value across following $10$ slots -- $\{4, 24, 44, \dots, 94\}$ and divide it by $\mathcal{K} = \frac{100}{20} = 5$ to find the average RSRP.
        
        %and $j$ is the subchannel index of the single-slot resource
        \item If the remaining slots in $S_A$ is less than $X\%$ of the total available slots (Step $6$), then $P_{th}$ is increased by $3$ dB (Step $7$), and Steps $3$ to $5$ are repeated. In NR-V2X Mode-2, X can be set to 20, 35 or 50.
        \item Each vehicle selects a random slot resource from the resources remaining in $S_A$ for transmission in Step $7$\footnote{Note that Release 14 SPS had an additional selection criteria that selected slot resources with the lowest sidelink received strength indicator (S-RSSI) measurements. This was removed in Release 16 to accommodate smaller RRIs \cite{Zoraze_NRV2X_Mode2_overview}. }. 
    \end{enumerate}

    % Else as shown in step 7, the set $S_A$ is sorted in the non-increasing order of observed RSRP and the algorithm initializes another set $S_B$ with the first $20\%$ of the candidate slots from set $S_A$. Finally in step 8, a random slot from $S_B$ is chosen as the chosen slot ($s^{*}$) from the considered transmitting vehicle.
    
    \item \textbf{Resource Reselection (Steps 8-10):} Each vehicle can reserve the same slot (selected in Step $7$) for next \textit{Resource Counter (RC)\footnote{Resource Counter (RC) is the maximum number of transmissions a certain vehicle is allowed (by utilizing the selected slot/resource in the current selection window) before having to reselect a new set of resources.}} number of subsequent transmissions with the same transmission interval, i.e., \textit{RRI}. %Notice from 3GPP mandated constraints in Table \ref{TABLE_SPS}, that the possible values of the RC depend on the RRI. 
    The RC varies with the RRI to ensure that the selected slot/resource is in use for at least $0.5$ s and at most $1.5$ s. This means that for a $20$ ms RRI, $25\leq \textnormal{RC} \leq 75$, for $50$ ms RRI, $10\leq \textnormal{RC} \leq 30$, and for a $100$ ms RRI, $5\leq \textnormal{RC} \leq 15$.

    \hspace{0.2in}After RC reaches $0$, the vehicle can either continue utilizing the preselected resources with a probability $p_r$ or reselect new resources for BSM transmissions with a probability $(1 - p_r)$ (See Steps $8$-$9$). 
\end{itemize}

\section{Ch-RRI SPS: Semi-persistent Scheduling Using Channel Aware RRI Selection}
\label{Section: Ch-RRI}
In this section, we present the limitations of conventional SPS protocol in terms of improving cooperative awareness performance of NR-V2X Mode-2, followed by detailed discussion on how to overcome them through a novel channel aware RRI selection algorithm (Ch-RRI SPS). The Ch-RRI SPS discussion first formulates the channel aware RRI selection problem as an Integer Linear Programming (ILP) problem and then goes into the algorithmic details of the proposed Ch-RRI SPS protocol.

% We present the details of Ch-RRI SPS, which is SPS powered by a channel aware RRI selection algorithm (Ch-RRI). Ch-RRI uses available channel resources to choose the lowest possible RRI without increasing packet loss due to congestion. We first formulate the channel aware RRI selection problem as an Integer Linear Programming (ILP) problem and followed by algorithmic details of the proposed Ch-RRI powered SPS protocol.

\subsection{Limitations of SPS on Cooperative Awareness}
 
 \begin{figure}[h]
\centering
\includegraphics[width=\columnwidth]{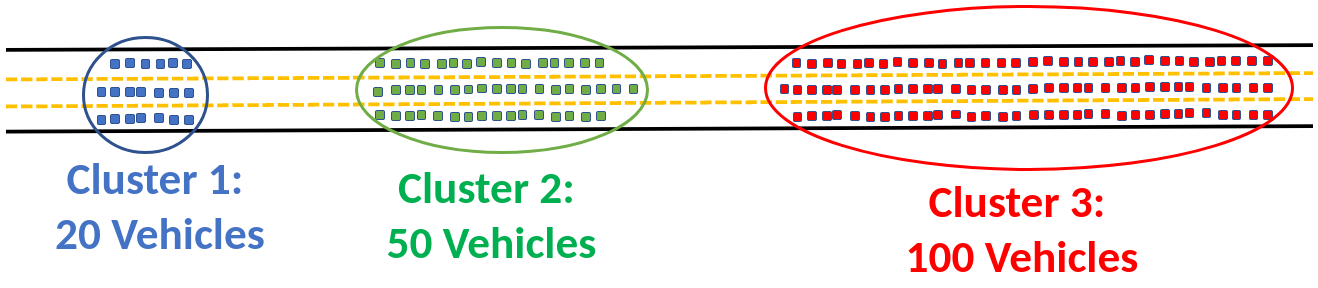}
\caption{NR-V2X example network with three clusters of vehicles.}
\label{cluster_fig}
\centering
\end{figure} 

We present the limitations of SPS through a simple NR-V2X example (See Fig. \ref{cluster_fig}). The example NR-V2X network consists of three clusters\footnote{Each vehicle is at 1-hop (i.e., within the transmission range) of every other vehicles belonging to a certain cluster of vehicle.}, of vehicles, where clusters $1$, $2$, and $3$ respectively have $20$, $50$, and $100$ vehicles. We make the following \textbf{assumptions} for all example NR-V2X scenarios.
\begin{itemize}
    \item We assume $T_1 = 0$ and $T_2 = $ RRI, which means, the size of selection window is equal to the RRI.
    \item The NR-V2X physical layer consists of $2$ subchannels only. Each BSM transmission uses both the subchannels and takes 1 ms to transmit. 
    %It means, if the selection window is $100$ ms (i.e., RRI = $100$ ms), there at most $100$ distinct vehicles has BSM transmission opportunities (with no potential collisions from other vehicles).
    This means if the selection window is $100$ ms (i.e., RRI = $100$ ms), then at most $100$ distinct vehicles have unique BSM transmission opportunities (assuming no collision in resource selection).

    % Thus, if RRI is 100 ms for each vehicle, then there are at most $100$ BSM transmission opportunities
    \item Each cluster of vehicles is sufficiently spaced apart from each other so that there is no inter-cluster interference. This means, for example, no transmissions from cluster $2$ interfere with any transmissions from cluster 1, and vice-versa.     
    
\end{itemize}

Under the above assumptions, let us look at the (i) \textit{Channel Occupancy Percentage $C_{occup}^{(i)}$}, defined as the percentage of the number of vehicles transmitting to the total number of available slots transmission opportunities for the $i^{th}$ cluster, and (ii) \textit{Probability of Successful Reception ($P_{suc})$}. For simplicity, let $P_{suc}$ is given by $\frac{1}{N}$ where $N$ is the number of vehicles using the same slot for BSM transmissions. Note that we only make this $P_{suc}$ assumption here to simplify this discussion on the limitations of fixed RRI SPS. In simulations and reality, $P_{suc}$ gets worse as $N$ increases. 

%\textbf{Under SPS with fixed RRI.} 

Table \ref{table:SPS_metrics} depicts the average $C_{occup}^{(i)}$ and $P_{suc}$ observed in the example ideal NR-V2X network (meaning slot resources are distributed evenly across all vehicles)  under conventional SPS with three different values of RRIs, i.e., $20$ ms, $50$ ms, and $100$ ms. For instance, the average $C_{occup}^{(i)}$ for SPS with low RRI = $20$ ms is given by $\frac{\sum RRI \times (C_{occup} ^{(i)}}{\text{total number of vehicles}} $. $C_{occup}^{(i)}$ for each cluster can be computed as follows:
Since RRI is $20$ ms, there are $20$ transmission opportunities (or slots), (i) in cluster 1, $20$ vehicles attempt to transmit, which implies $C^{(1)}_{occup} = \frac{20}{20} \times 100 = 100\%$, (ii) in cluster 2, $50$ vehicles attempt to transmit, which results in $C^{(2)}_{occup} = \frac{50}{20} \times 100 = 250\%$, and (iii) in cluster 3, $100$ vehicles attempt to transmit, resulting in $C^{(3)}_{occup} = 500\%$. Thus, the average $C_{occup}$ is $\frac{20 \times C^{(1)}_{occup} + 50 \times C^{(2)}_{occup} + 100 \times C^{(3)}_{occup}}{20 + 50 + 100} \times 100  = 379.4\%$. The average $P_{suc}$ can be computed in the similar fashion, and it turns out to be $0.35$ in case of SPS with RRI as $20$ ms. On contrary for SPS with high RRI = $100$ ms  the average $C_{occup}$ and $P_{suc}$ are $75.88\%$ and $1$ respectively.

 \begin{table}[]
    \centering
     \caption{Ideal Channel Occupancy Percentage and Success Probability of Conventional SPS with Fixed RRIs vs Scheduling with Adaptive RRI}   
    \label{table:SPS_metrics}
    \begin{tabular}{|p{2.5cm}|p{0.75 cm}|p{0.75 cm}|p{0.75 cm}|p{1cm}|p{1.8cm}|}
    \hline
     Metrics & \multicolumn{3}{c}{Conventional SPS} & \textbf{Ch-RRI}  \\ \hline
      & 20 ms & 50 ms & 100 ms & \textbf{Adaptive RRI}\\ \hline
     Channel Occupancy Percentage & 379.4\% & 152.94\% & 75.88\% & 100\% \\ \hline
     Probability of Successful Reception & 0.35 & 0.7058 & 1 & 1 \\ \hline
    \end{tabular}
    \vspace{-0.1in}
\end{table}
 
Note that SPS with low RRI such as, $20$ ms leads to \textit{overly congested radio channels} ($379.4\%)$, and thus, large number of dropped BSM packets ($0.35$), particularly, in clusters $2$ and $3$ with $>20$ vehicles). The lost packets result in high tracking error, which compromises the awareness of considered NR-V2X network. Whereas, in case of SPS with high RRI as $100$ ms, the radio resources are under-utilized ($75\%$), particularly in cluster $1$ and $2$ with $<100$ vehicles. Tracking error performance can be significantly improved by choosing lower value of RRI as lower value of RRIs will improve timely delivery of BSMs. From the above discussion, it is evident that SPS with fixed RRI (irrespective of the chosen value of RRI) is limited in the context of improving overall cooperative awareness of NR-V2X networks. 

%\textbf{Under Ch-RRI scheduling with adaptive RRI.} 

To better address the limitations of static RRI SPS, we propose Ch-RRI, an adaptive RRI selection algorithm where each vehicle chooses its RRI based on the neighborhood density. Ideally, in the example NR-V2X network depicted in Fig. \ref{cluster_fig}, Ch-RRI would choose RRI = $20$ ms for a vehicle in cluster $1$ (with $20$ vehicles). Similarly, Ch-RRI will choose RRI = $50$ ms for cluster $2$ (with $50$ vehicles) and RRI = $100$ ms for cluster $3$ (with $100$ vehicles) -- which will result in $C_{occup} = 100\%$ and $P_{s} = 1$ (See Table \ref{table:SPS_metrics}). It means that the proposed Ch-RRI strategy with adaptive RRI enables judicious utilization of the radio resources. This in turn reduces the tracking error and enhances the cooperative awareness of NR-V2X networks.  
% \vspace{-.2 in}
\subsection{Problem Formulation}

\textbf{Notations.} At each time instant $t \in T$, let $s(t,P_{th})$ denote the set of available slots (slots with observed power less than $P_{th}$) in the neighborhood of any vehicle $v$. RRI$_v$ is the $v^{th}$ vehicle's RRI and $\mathcal{N}$ is the set of all vehicles in the environment.%, and $\mathcal{N}_v \subseteq \mathcal{N}$ is the set of all neighboring vehicles of $v$.

\textbf{Objective function.} As shown in Eq. \ref{eq:optProb}, the objective function is to minimize the resource reservation interval (RRI) (or maximize BSM rate) at each vehicle $v$ (where $v \in \mathcal{N}$) over the entire time duration $T$.

\begin{align} %\label{problem_formulation}
	% &\!\min_{r_{v}(t)} \sum_{t \in T} \frac{1}{\mathcal{N}(\mathcal{N}-1)}  \sum_{v \in \mathcal{N}} \sum_{u \in \mathcal{N}, u \neq v}  TAoI_{uv}(t) \label{eq:optProb}\\
	&\!\min \sum_{t \in T} \frac{1}{|\mathcal{N}|}  \sum_{v \in \mathcal{N}}  \textnormal{RRI}_{v}(t) \label{eq:optProb}\\
	&\text{subject to} \sum_{v \in \mathcal{N}}{\textnormal{RRI}_{v}(t) \leq s(t,P_{th})},  \forall v \in \mathcal{N}, t \in T \label{eq:constraint1}\\
	%&\text{where }  s(t)=\\
	%&P_{v} \leq P_{th}, \forall v \in V \\
	&  \textnormal{RRI}_{min} \leq \textnormal{RRI}_{v}(t) \leq \textnormal{RRI}_{max}, \forall v \in \mathcal{N}, t \in T \label{eq:constraint2}
\end{align}

% Ch-RRI can be formulated as shown in Eqs. \ref{eq:optProb} - \ref{eq:constraint2}.

\textbf{Constraints.} Eq. \ref{eq:constraint1} constrains that the sum total of the RRI interval must be less than the number of slots available to each vehicle $v$ at time instant $t$. This constraint ensures that the channel congestion or the outage probability across all vehicles are not compromised.  Eq. \ref{eq:constraint2} restricts the RRI for each vehicle to be within the range $[$RRI$_{min}, $RRI$_{max}]$. 

Note that solving the aforestated ILP formulation  would provide the optimal solution to Ch-RRI. However, this is impractical in a real-world NR-V2X setting because of the dynamic nature of the setup (mobility of vehicles) and lack of global knowledge. Additionally, in order to solve the ILP problem, we would require the knowledge of available slots at future time instants, which is impractical to obtain in time varying NR-V2X networks. Therefore, our proposed Ch-RRI SPS is based on decentralized SPS and estimates the latest channel occupancy (via sensing window), selecting the suitable value of RRI at each vehicle in the NR-V2X network based on its local knowledge.
%Thus, our proposed Ch-RRI SPS protocol, similar to SPS, is designed as a decentralized protocol that estimates the latest channel occupancy (via sensing window) and selects the suitable value of RRI at each vehicle in the NR-V2X network based on its local knowledge.

\begin{figure}[h]
\centering
\includegraphics[width=\columnwidth]{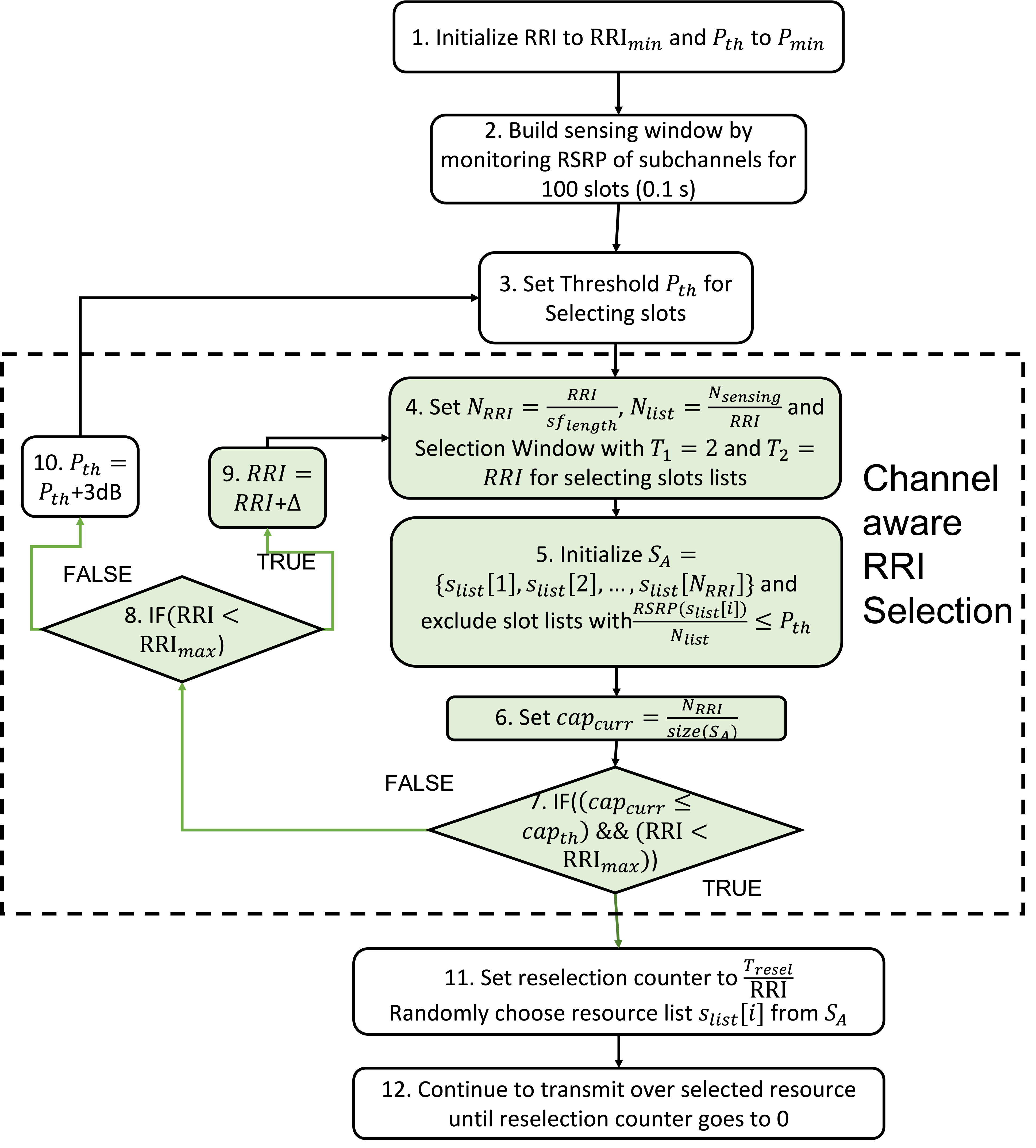}
\caption{Flowchart of Ch-RRI SPS.}
\label{adaptive_sps_algorithm_fig}
\centering
\vspace{-0.2 in}
\end{figure}

\subsection{Ch-RRI SPS Description}
%\Avik{Remove steps 4, and 8 and try to explain/focus on differences in Ch-RRI algorithm from sps}
This subsection discusses in detail the proposed Ch-RRI SPS protocol. As shown in Fig. \ref{adaptive_sps_algorithm_fig}, Ch-RRI SPS makes significant enhancements to the conventional SPS algorithm. The green boxes represent the steps of the Ch-RRI algorithm and all other steps are from the conventional SPS. %Next, we explain Ch-RRI algorithm as discussed below.

% In this section, we present the Ch-RRI algorithm. The goal of Ch-RRI is to transmit with the lowest possible RRI in between BSMs, without interfering with another vehicles transmissions. If there are many other vehicles using the channel, and not enough available slots under the lowest possible RRI, the vehicle selecting resources searches for available slots with a larger RRI between transmissions. This procedure ensures that vehicles transmit with the smallest possible RRI between messages without compromising on packet losses due to channel congestion. 

% \par To accomplish this, Ch-RRI uses the sensing procedure from SPS to measure interference on the channel and builds slot \textit{lists} under possible RRIs. A slot list is a set of possible slot resources spaced out by the RRICh-RRI compares the average RSRP (similar to SPS) of each slot list to $P_{th}$, and those lists with a RSRP larger than $P_{th}$ are excluded. %We explain the Ch-RRI algorithm, outlined in Fig. \ref{adaptive_sps_algorithm_fig}, below.
%Fig. \ref{adaptive_sps_algorithm_fig} shows the Ch-RRI algorithm, with red outlined boxes detailing steps used from the original SPS algorithm. 
  
\begin{itemize}
    %\item \textbf{RRI Initialization and Sensing (Steps 1-2):} Unlike SPS, Ch-RRI initializes estimated RRI to $RRI_{min}$, besides initializing the RSRP threshold ($P_{th}$) to a minimum value $P_{min}$ (see Step 1). Following this, Ch-RRI continuously monitors the slots by measuring RSRP and stores the sensing measurements for the sensing window, i.e., $N_{\rm sensing}$ slots (Step 2). We refer to slots as s $N_{SW}$
    
    \item \textbf{RRI Initialization and Sensing (Steps $1$-$2$):} Similar to SPS, Ch-RRI SPS continuously measures the RSRP and S-RSSI of the previous $N_{\rm sensing}$ slots and stores the sensing measurements in the sensing window (Step $1$). In Step 2, Ch-RRI initializes the estimated $\widehat{\textnormal{RRI}}$ to $\textnormal{RRI}_{min}$ and the the RSRP threshold ($P_{th}$) to a minimum value $P_{min}$. The selection window in Ch-RRI SPS is not initialized before the resource selection process has started (see Step $2$ in Fig. \ref{SPS_algorithm_fig}), and as available resources are identified and the estimated $\widehat{\textnormal{RRI}}$ is changed, the selection window is updated. As in SPS, Ch-RRI SPS sets and updates the $P_{th}$ (See Step 3).

    \item \textbf{Ch-RRI: Channel-aware RRI Selection (Steps 4-10):}  Each vehicle utilizes $N_{\rm sensing}$ slots (obtained in Step 2) for identifying the available slots and subsequently, selecting the minimum RRI possible in between transmission while ensuring that there remain resources (or slots) for other vehicles. %A set containing $N_{RRI}$(the size of the ) slot lists ($s_{list}$) are created for every considered RRI. Each slot list coins the slots spaced out by the RRI. So for example, with an RRI of 10, the first slot list $s_{list}[1]$ would consist of ($s_1$, $s_{11}$, $s_{21}$,$\dots$,$s_{N_RRI}$). $N_{list}$ refers to the number of slot lists possible under the considered RRI and $s_{length}$ is the length of the slot in ms. Explain 5-6 well 
    \begin{enumerate}
        \item  In Step 4, Ch-RRI updates the estimated $\widehat{\textnormal{RRI}}$ and initializes the selection window with $T_1=1$ and $T_2=\widehat{\textnormal{RRI}}$. $T_1$ is fixed to $1$ to maximize slot resources. 
        
        \item Each vehicle populates set $S_A$ with all slots in the selection window and $S_B$ as an empty set (See Step 5). The candidate slot exclusion criteria is also borrowed from SPS, except that, $\widehat{RRI}$ is an adjustable parameter in Ch-RRI.
        
        %\item Each vehicle excludes single-slot resources from $S_A$ if-- (i) the vehicle has not monitored the corresponding candidate slot in the sensing window and (ii) the RSRP measurement for corresponding candidate slot is higher than $P_{th}$. The RSRP exclusion criteria for the $i^{th}$ slot in the selection window for Ch-RRI can be rewritten as:

        % \begin{equation} \label{RSRP_calculation1}
        % \hspace*{-1.5cm}     \frac{1}{\frac{N_{\rm sensing}}{\widehat{\textnormal{RRI}}}}RSRP\sum_{k=0}^{\frac{N_{\rm sensing}}{\widehat{\textnormal{RRI}}}}\left(s_{n+T_1+i-N_{\rm sensing}+k \cdot \widehat{\textnormal{RRI}}}^{j}\right)\leq P_{th}
        % \end{equation} 
        
        %$RSRP(\sum_{k=0}^{}s_{n+T_1+i-N_{\rm sensing}+k \cdot RRI}^{j})$
        \item If the number remaining slots in $S_A$ is less than $20\%$ of the total available slots (Step $7$), then, Ch-RRI, unlike SPS, first checks whether the $\widehat{\textnormal{RRI}} < \textnormal{RRI}_{max}$ (Step $8$). If yes, $\widehat{\textnormal{RRI}}$ is increased by $\Delta$ as shown in Step $9$, and Steps $4$ - $7$ are repeated. Once $\widehat{\textnormal{RRI}}$ has reached $\textnormal{RRI}_{max}$, then $P_{th}$ is increased by $3$ dB in Step 10, and Steps $3$ - $7$ are repeated.
        
        %but $\widehat{\textnormal{RRI}}< \textnormal{RRI}_{max}$, this indicates that the slot resources in $\textnormal{RRI}_{min}$ will not support all of the vehicles transmitting, but $\widehat{\textnormal{RRI}}$ can still be expanded. Therefore $\widehat{\textnormal{RRI}}$ is increased by $\Delta$ (Steps 8-9) and Steps 4-7 are repeated. Once $\widehat{\textnormal{RRI}}$ is maximized, the $P_{th}$ is increased by $3$ dB (Step 7), and Steps $(3-6)$ are repeated. This step is not present in SPS.
        
        %\item \Vijay{If the ratio of occupied to the total number of slot lists is high, determined by $cap_{th}$,} this indicates that the current RRI is too low, and must be increased (step 11). If the RRI is at $RRI_{max}$, then similar to SPS, the $P_{th}$ threshold is increased by 3 dB (step 12). 
    \end{enumerate}

    \item \textbf{Resource Selection} (Step $11$) and \textbf{Resource Reselection} (Step $12$) are similar to SPS. As in SPS, a reselection counter (RC) value is chosen such that the resource reservation is restricted between $0.5$ and $1.5$ s, irrespective of chosen $\widehat{\textnormal{RRI}}$. However, in the case of Ch-RRI SPS, since RC is zero, unlike SPS (see Steps $8$-$9$ in SPS flowchart), Ch-RRI SPS does not allow re-reservation of slot resources. Both these modifications are to ensure that Ch-RRI allows each vehicle to adjust its RRI at the time of resource reselection and account for changing vehicle traffic conditions.

    %Each vehicle can reserve the same slot (selected in Step 12) for next RC number of subsequent transmissions with the same selected transmission interval, $\widehat{\textnormal{RRI}}$. As in SPS, RC varies with the RRI to ensure that the selected slot/resource is in use for at least 0.5 s and at most 1.5 s.
    
    %In step 11, if the number of avaiable slot lists are sufficient under the current RRI, the current RRI and a random slot list from $S_A$ is chosen. The reselection counter determines how many times each vehicle is allowed to transmit over the chosen slot and RRI. In Ch-RRI, the reselection counter is set to $\frac{500 ms}{RRI}$ where the RRI is in ms. This ensures that each vehicle transmits over the chosen resources for approximately 0.5 s, similar to SPS.  
\end{itemize}

\section{AoI-RRI: Age of Information Aware RRI Selection Algorithm}
\label{Section: AoI aware SPS}
%\Vijay{Would you like to give some insights for AoI-RRI selection algorithm on top of Ch-RRI algorithm?} 
In this section, we provide some intuitions on the potential limitations of Ch-RRI, provide a brief background of Age-of-Information (AoI), and formulate the problem of AoI minimization in vehicular networks as an ILP problem. We discuss why the ILP formulation is impractical to solve for a decentralized network and propose AoI-RRI, a decentralized AoI-aware RRI selection algorithm. 

\subsection{Limitation of Ch-RRI on Cooperative Awareness}
Though Ch-RRI is a promising RRI adaptation algorithm and promises to improves the cooperative awareness as discussed in Section \ref{Section: Ch-RRI}, it has certain limitations as presented in this section. 
\begin{itemize}
    \item Ch-RRI SPS is based on the accurate knowledge of channel occupancy at any given time, however, given the dynamics of vehicles and channel usage, obtaining accurate channel availability information at each vehicle is not possible.
    \item Ch-RRI SPS attempts to enable each vehicle to transmit at the minimum RRI while accounting for 100\% channel occupancy. However, the optimal trade-off between channel occupancy and optimal RRI may not be at 100 \% channel occupancy.   
\end{itemize}
Motivated by these limitations, we consider optimizing for a novel information freshness metric, namely AoI. Since optimizing for AoI guarantees freshness (and thus the best tracking error in this context), we can better account for trade-off between the optimal RRI and channel occupancy.
\subsection{Age of Information}
\label{Subsection:Age of Information}
\begin{figure}[h]
\centering
\includegraphics[width=\columnwidth]{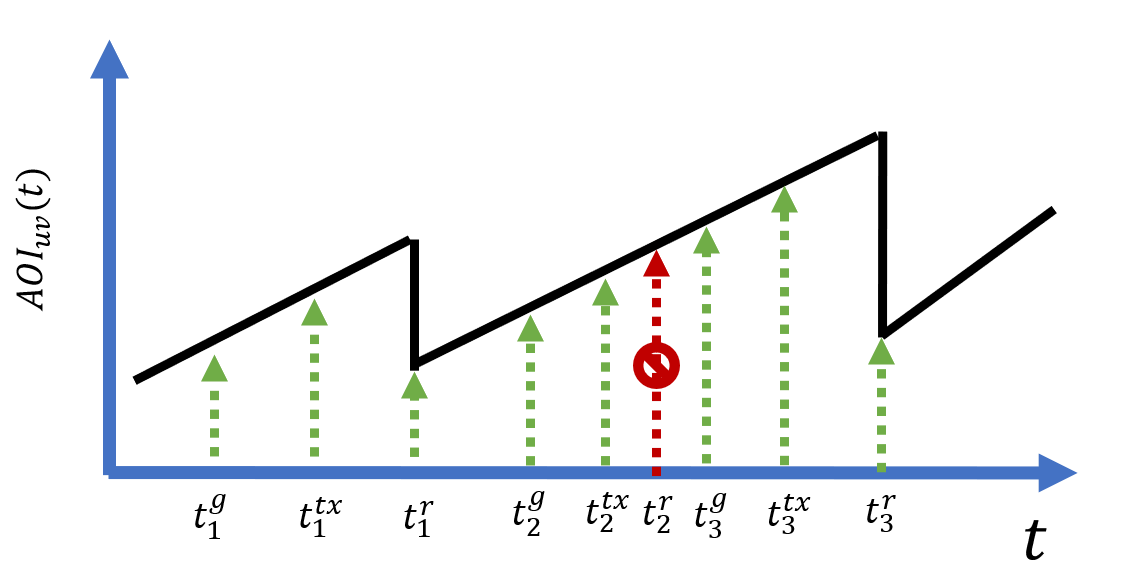}
\caption{$AoI_{uv}$ is the AoI at vehicle $v$ based on the latest received transmission from vehicle $u$.}
\label{AOI_illustr_fig}
\centering
\end{figure}

The AoI quantifies the \textit{freshness} of information at any receiving node that was originally generated and transmitted by the transmitting node\cite{kaul2012real}. 
For the purposes of vehicular networks, AoI is the elapsed time since the last received location packet was generated at the transmitting vehicle. 
%In the context of V2V networks, AoI is the time elapsed since the last successfully received location beacon at the receiving vehicle that was generated at the sender vehicle. 
Let $t^g$ denote the time that the packet containing the most recent location at sender vehicle $u$ was generated. Assuming a linear cost function for the AoI evolution, the AoI at the receiving vehicle $v$ assuming the current time is $t$ is:
\begin{equation} \label{AoI_calculation}
    \textnormal{AoI}_{uv}(t) = t - t^g.
\end{equation}
The evolution of the AoI for vehicle $v$ is illustrated in Fig. \ref{AOI_illustr_fig}. The generation, transmission, and reception times of the $i^{th}$ location packet are denoted by $t^{g}_{i}$, $t^{tx}_{i}$, and $t^{r}_{i}$. Therefore, the overall delay for a packet = $(t^{r}_{i} - t^{g}_{i})$. %Assuming the ${(i+1)}^{th}$ packet is generated at $t^{g}_{(i+1)}$, the generation rate is given by $\Delta = \frac{1}{t^{g}_{i+1} - t^{g}_{i} }$. 
% Fig. \ref{AOI_illustr_fig} illustrates AoI$_{uv}(t)$, the AoI at the receiving vehicle $v$ after a sender vehicle $u$ generates and transmits a message, as a function of time.

Eq. \ref{AoI_calculation} can be used to find the average AoI$_{uv}$ between vehicle $v$ and neighboring vehicle $u$. The average AoI$_{uv}$ is found by calculating the area under AoI$_{uv}(t)$. The area under AoI$_{uv}(t)$ is normalized by the observation time $T_{obs}$ \cite{kaul2012real}  
%$ AoI_{uv} = \frac{1}{T} \int_{T} AoI_{uv}(t)$. 

\begin{equation}
\label{eqn:pairwise_Avg_AoI}
    \textnormal{AoI}_{uv} = \frac{1}{T_{obs}} \int_{T_{obs}} \textnormal{AoI}_{uv}(t).
\end{equation}
We represent the set of all vehicles in the environment as $\mathcal{N}$ and the set of all neighboring vehicles of $v$ as $\mathcal{N}_v \subseteq \mathcal{N}$, where $u \in \mathcal{N}_v$. The average AoI at vehicle $v$ is:

\begin{equation}
\label{eqn:neigh_Avg_AoI}
    \textnormal{AoI}_v = \frac{1}{|\mathcal{N}_{v}|} \sum_{u \in \mathcal{N}_{v}} \textnormal{AoI}_{uv}.
\end{equation}
The overall system AoI with $|\mathcal{N}|$ vehicles can be computed as:

\begin{equation}
\label{eqn:system_AoI}
    % AoI = \frac{1}{\mathcal{N}(\mathcal{N}-1)} \sum \limits_{u \in \mathcal{N}} \sum \limits_{v \ne u, v \in \mathcal{N}} AoI_{uv}
    \textnormal{AoI} = \frac{1}{|\mathcal{N}||(\mathcal{N}-1)|} \sum \limits_{v \in \mathcal{N}} \sum \limits_{u \in \mathcal{N}_v} \textnormal{AoI}_{uv},
\end{equation}
where $|\mathcal{N}||(\mathcal{N}-1)|$ are the number of unique pairs of sender and receivers in the system. 
%\subsection{V2V Assumptions for AOI}
Assuming the simple vehicular network in Fig. \ref{tracking_error_fig} with vehicles $u$ and $v$, we can assume that inter-packet reception interval between receiver $v$ and transmitter $u$ are multiples of the RRI. Using these assumptions and the fact that the AoI resets to 0 upon successful reception of location information, the evolution of the AoI at the receiving vehicle can be written as

\begin{equation} \label{AoI_slot}
    \textnormal{AoI}(t + \textnormal{RRI}) =
    \begin{cases}
     0, & \text{if} \hspace{0.04in}  \text{a packet is received}  \\
     \textnormal{AoI}(t)+\textnormal{RRI}, & \text{otherwise}
    \end{cases}
\end{equation}
Note that Eq. \ref{AoI_slot} implies that minimizing the RRI also minimizes the AoI at the receiving vehicle. However, too many vehicles transmitting with a small RRI can increase the number of lost packets due to congestion, which also increases the AoI. Therefore, as in the case of tracking error, there is a tradeoff between smaller RRIs and improved AoI performance.  

\subsection{AoI ILP Problem Formulation}
%For this paper, we consider a out of coverage highway scenario where $N$ vehicles share $B$ RBs over a bandwidth $W$, and every vehicle $i$ is equipped with a radio using NR-V2X Mode-2. The $i^{th}$ vehicle is allowed to schedule a packet at time $t$, and subsequent transmissions at intervals of RRI$_v$. 
The objective of AoI-RRI is to minimize the overall system AoI based on what each vehicle observes. As in the Ch-RRI problem formulation, finding an optimal RRI that minimizes the system AoI can be written as a Integer Linear Programming (ILP) problem. % Each vehicle in the environment is given a different priority level($p$), from $1$to 5, which 5 being the highest priority packet. The function $f(r_{v}(t))$ determines the priority level for each vehicle, and maps the $v^{th}$ vehicles priority to a AoI threshold for that priority($AoI_{th_{uv}}$), with highest priority vehicles given the lowest AoI threshold. Based on each vehicles priorities and tracking error req. The $v^{th}$ vehicle is allowed to schedule a packet at time $t$, and subsequent transmissions at intervals of RRI$_v$. We try to minimize the tracking error as a function of each vehicles priority.

%\textbf{Notations.} Let $sf(t,P_{th})$ denote the set of available slots (slots with observed power less than $P_{th}$) in the neighborhood of any vehicle $v$ at each time instant $t \in T$. RRI$_v$ is the $v^{th}$ vehicle's RRI, $C(t)$ is the total channel capacity in the NR-V2X network, and $\mathcal{N}$ is the number of vehicles in the neighborhood (i.e., within transmission range) of vehicle $v$. Assume that $v$ is transmitting to vehicle $u$. 

\textbf{Objective function.} Eq. \ref{eq:AOI_optProb} is the objective function where each vehicle chooses a RRI at each vehicle $v$ (where $v \in \mathcal{N}$) such that the AoI is minimized over the entire time duration $T$. % As shown in Eq. \ref{eq:optProb}, the objective function is to minimize the resource reservation interval (RRI) (or maximize BSM rate) at each vehicle $v$ (where $v \in \mathcal{N} \cup v$) over the entire time duration $T$. 

\begin{align} %\label{AoI_problem_formulation}
    %&\!\min_{RRI_v} \sum_{t \in T} \frac{1}{\mathcal{N}}  \sum_{v \in \mathcal{N}} w_v e^{track}_v(t) \label{eq:optProb}\\
    &\!\sum_{t \in T} \frac{1}{|\mathcal{N}|}  \sum_{u, v \in \mathcal{N}}  \textnormal{AoI}_{uv}(t). \label{eq:AOI_optProb}\\
    %&\text{subject to} \sum_{u, v \in \mathcal{N} }{w_v \mathbb{E}[AoI_{uv}(t)] \leq f(r_{v}(t))},  \forall v \in \mathcal{N}, t \in T \label{eq:constraint1}\\
    %&\text{subject to} \sum_{u, v \in \mathcal{N} }{\frac{1}{\textnormal{RRI}_v} \leq C(t)},  \forall v \in \mathcal{N}, t \in T \label{eq:AOI_constraint1}\\
    %&\text{subject to} \sum_{v_{hp} \in \mathcal{N} \cup v}{RRI_{v}(t) \leq s(t,P_{th})},  \forall v \in \mathcal{N}, t \in T \label{eq:constraint1}\\
    %&\text{subject to} \sum_{v \in \mathcal{N_{hp}} \cup v}{s_v RRI_{v}(t) \leq RRI\_func(TTC,t,P_{th})},  \forall v \in \mathcal{N}, t \in T \\v_s \mathbb{E}[T_\textnormal{IRT}] \\
    %&\text{where } \mathbb{E}[AoI_{uv}]= \left( \frac{\textnormal{RRI}_v}{P_s}\right)
    %\label{eq:constraint1}\\
    %&\text{where }  s(t)=\\
    &\text{subject to }  \textnormal{RRI}_{min} \leq \textnormal{RRI}_{v}(t) \leq \textnormal{RRI}_{max}, \forall v \in \mathcal{N}, t \in T \label{eq:AOI_constraint2}
\end{align}

% Ch-RRI can be formulated as shown in Eqs. \ref{eq:optProb} - \ref{eq:constraint2}.

\textbf{Constraints.} Eq. \ref{eq:AOI_constraint2} restricts the RRI for each vehicle to the range $[$RRI$_{min}, $RRI$_{max}]$. Note that there are no capacity constraints for Eq. \ref{eq:AOI_optProb}, and the AoI minimization itself decides the optimal tradeoff. 

%Although the above ILP formulation provides a centralized optimal solution to the AoI minimization solution, it is considered impractical because the global information is not available at every vehicle. The exact computation of AoI for vehicle $v$ would require knowledge of a packets probability of success ($p_s$) for all neighboring vehicles, which can at best be estimated based on transmissions from those neighboring vehicles. 
However, as in the case of Eq. \ref{eq:optProb}, solving Eq. \ref{eq:AOI_optProb} assumes global information at every vehicle and knowledge of each vehicles channel information and mobility even at future time instants, which is not realistic. As a result, this ILP is not practical. 
Thus we propose AoI-RRI, an algorithm that attempts to iteratively minimize the NR-V2X system AoI in a decentralized manner, described in Section \ref{Subsection: AoI aware SPS algorithm}

\subsection{AoI-RRI SPS}
\label{Subsection: AoI aware SPS algorithm}
This section details AoI-RRI SPS, SPS powered by a decentralized AoI-aware RRI selection algorithm. AoI-RRI finds a RRI$_v$ (where RRI$_v\in[\textnormal{RRI}_{min}$  $\textnormal{RRI}_{max}]$) for vehicle $v$ by iterating the previously chosen RRI, denoted RRI$^{t-1}$, such that the locally measured average AoI is minimized. The algorithm runs in three steps. First, at the RRI selection time, $v$ uses the sensing window history to measure the channel congestion in the network. Second, $v$ calculates the time since the last received packet from every vehicle in the vicinity to estimate the local average AoI, and adjusts its RRI in order to minimize this last measured AoI. Finally, $v$ uses the semi-persistent scheduling procedure presented in Fig. \ref{SPS_algorithm_fig} and selects slot resources using the latest selected RRI. The algorithm details are as follows: 

%\textbf{Step 1: Sensing and AoI estimation}
\textbf{Step 1: Sensing and Channel Congestion Detection.} 
%As shown in Algorithm 1, vehicle $v$ uses Eqs. \ref{AoI_calculation} and \ref{eqn:neigh_Avg_AoI} and calculates its local AoI based on transmissions received by neighboring vehicles. The AoI is computed based on when the packet was last received.
In line \ref{algo_1_cong} of Algorithm $1$, AoI-RRI uses a channel congestion detection algorithm, further detailed in Algorithm $2$, to measure and detect whether or not significant channel congestion has occurred. The channel congestion algorithm returns \textbf{true} if there is significant congestion detected and like Ch-RRI SPS, uses the previously chosen $\textnormal{RRI}^{t-1}$ and the RSRP sensing measurements for the last $N_{\rm sensing}$ slots, i.e. the sensing window, as inputs. In line $2$ of Algorithm $2$, $v$ calculates the size of the set $S_{\rm total}$, which is the set of all possible slots resources using RRI$^{t-1}$. Lines 3-5 of Algorithm $2$ populate set $S_{\textnormal{avail}}$ with those slots with linear average RSRP measurements higher than $P_{th}^{t-1}$. If the number of slots in $S_{\rm avail}$ is less than $20\%$ of the total available slots, it means that network congestion has increased since the previous resource reservation, and RRI$^{t-1}$ or $P_{th}^{t-1}$ must be increased. Using RSRP measurements can give a robust estimate of the channel congestion without adding any sensing overhead since the sensing window measurements and RSRP threshold are taken as a part of the SPS procedure. If the channel congestion increases, lines $2$-$3$ of Algorithm 1 automatically increase the RRI to avoid congesting the channel further.  
 %\textbf{Step 2: RRI adjustment}: 
\par \textbf{Step 2: AoI Estimation and RRI Adjustment.} 
If the channel has not become significantly congested, vehicle $v$ uses Eqs. \ref{AoI_calculation} and \ref{eqn:neigh_Avg_AoI} and computes its local AoI by calculating and averaging the time since the last received packet from each neighboring vehicle. The AoI-RRI algorithm iterates upon the last chosen RRI to minimize the most recent average AoI at each vehicle $v$. AoI-RRI uses one of the following three actions-- (1) \textbf{DECR} - decrease RRI, (2) \textbf{INCR} - increase RRI, and (3) \textbf{SAME} - maintain the same RRI \footnote{The concept of utilizing these actions for adjusting the RRI is inspired by the algorithm presented in \cite{kaul2011minimizing}.}. Algorithm $1$ presents the pseudocode for this step.
%\par As shown in line \ref{algo_1_cong}, every vehicle $v$ starts by measuring the channel congestion of the network. AoI-RRI uses the $s_{\rm tot}$ RSRP sensing measurements for the last $N_{\rm sensing}$ slots, i.e. the sensing window, to detect whether or significant congestion has occurred. 

\par Lines $4$-$9$ form the core of AoI-RRI where $AoI_{avg}$ for the current and previous $T_{obs}$ are compared. In lines $4$-$6$ AoI-RRI checks if the local AoI has increased by a factor of more than $\sigma_t$, in which case the AoI has significantly worsened and AoI-RRI reverses the previously selected action. If in lines $7$-$9$, $v$ sees that the local AoI has decreased by a factor of more than $\sigma_t$, the previous action selected by the algorithm is repeated, i.e. the RRI continues to increase or decrease. If there is no significant change in AoI, AoI-RRI chooses the same RRI as before. Finally, in $10$-$20$, the new RRI is calculated based on the action selected in lines $2$-$9$. The new RRI returned is maintained until the end of the current $T_{obs}$. Further, $\beta$ decides the magnitude by which the RRI changes if the chosen action was \textbf{INCR} or \textbf{DECR}.

\begin{algorithm}[h]
\small
\small
\caption{AoI-RRI algorithm at vehicle $v$.}
%\
\textbf{Input:} RSRP of $N_{\rm sensing}$ slots across $J$ subchannels in sensing history, $s_{\rm total}=[s^{j}_1, \cdots ,s^{j}_{N_{\rm sensing}}]$, Minimum Power threshold ($P_{min}$), Maximum Power threshold ($P_{max}$), RRI chosen during the previous $T_{obs}$, $\textnormal{RRI}^{t-1}$, local average $AoI$ of the previous $T_{obs}$, $AoI_{avg}^{t-1}$, Action chosen during the previous $T_{obs}$, $\alpha^{t-1})$, Power threshold chosen during the previous $T_{obs}$, ($P_{th}^{t-1}$)\\
\textbf{Output:} New $\textnormal{RRI}^{t}$, slot $s_v$ for node $v$, and number of retransmissions $N_{resel}$
\begin{algorithmic}[1]

%\State{$\widehat{\textnormal{RRI}_v} = \frac{1}{\beta} \textnormal{ RRI}_{v}$} \Comment{New Potential RRI}
%\State{$P_{th} = P_{min}$} \Comment{Intializing threshold to minimum}

\State{$[\textnormal{channel}_{\textnormal{flag}}]= \textnormal{channel}_{\rm cong}(\textnormal{RRI}^{t-1}, s_{\rm total},P_{th}^{t-1}$)} \Comment{Checks to see if channel is congested}\label{algo_1_cong}

\If{$\textnormal{channel}_{\textnormal{flag}}==$true}
%\If{$AoI_{avg}^{t-1}>2\cdot \textnormal{RRI}_{avg}^{t-1}$}
    \State  {$\boldsymbol{\alpha^t}=$INCR  }
\EndIf    

    \If {$AoI_{avg}^{t} > AoI_{avg}^{t-1}+\sigma_t$ }
    \State  {$\boldsymbol{\alpha^t}=\textnormal{inverse}(\alpha^{t-1})$  } \Comment{Previous action caused AoI to worsen significantly, so inverse action}
    
    \ElsIf{$AoI_{avg}^{t} < AoI_{avg}^{t-1}-\sigma_t$ }
    \State  {$\boldsymbol{\alpha^t}=(\alpha^{t-1})$  } \Comment{Previous action caused AoI to get better so continue action}
    \Else
    \State  {$\boldsymbol{\alpha^t}=$SAME  } \Comment{AoI has not significantly changed from previous value, so maintain same RRI}
    \EndIf

%\Else
%    \State  {$\boldsymbol{\alpha_v} = {\alpha_v}$ } \label{algo_2:3_end} %\textit{/* increase interval as congestion detected*/}} 

\If {$\boldsymbol{\alpha^{t}}$ == INCR} \label{algo_2:7_start}
        \If{$\textnormal{RRI}^{t-1}\geq \textnormal{RRI}_{max}$}
        \State{$\textnormal{RRI}^t = \textnormal{ RRI}^{t-1}$}\Comment{AoI has improved but RRI has hit maximum}
        \Else
            \State{$\textnormal{RRI}^t =\beta \textnormal{ RRI}^{t-1}$}
        \EndIf
    \ElsIf {$\boldsymbol{\alpha^{t}}$ == DECR}
        \If{$\textnormal{RRI}^{t-1}\leq \textnormal{RRI}_{min}$}
            \State{$\textnormal{RRI}^t = \textnormal{ RRI}^{t-1}$} \Comment{AoI has improve but RRI has hit minimum}
        \Else
        \State{$\textnormal{RRI}^t =\frac{1}{\beta} \textnormal{ RRI}^{t-1}$}
        \EndIf
        
    \ElsIf {$\boldsymbol{\alpha^{t}}$ == SAME}
        \State{$\textnormal{RRI}^{t} = \textnormal{ RRI}^{t-1}$}
        %\State{$\Delta_v = \Delta'_{v}$} \label{algo_2:7_end}
        
\EndIf
\State {return $P_{th}$, $\textnormal{RRI}^{t}$}
%\State{$[s_i^{j}, N_{resel}, P_{th}]= s_{select}(\textnormal{RRI}^t, s_{\rm total})$} \Comment{Returns Optimal slot and number of retransmissions for required RRI}
\end{algorithmic}
\end{algorithm}
\textbf{Step 3: Semi-persistent Scheduling.} Finally, $v$ uses the RRI selected from Algorithm $1$ to choose a slot for transmission using the SPS procedure presented in Fig. \ref{SPS_algorithm_fig}. Note that the $P_{th}$ value that is used to select the slots is stored as a part of the procedure, and again the next time resources are selected. 

\begin{algorithm}[h]
\small
\caption{Channel Congestion}
%\
\textbf{Input:} All slots in sensing history, $s_{total} (1:N_{\rm sensing})$, Power threshold selected from previous slot selection ($P_{th}^{t-1}$), RRI selected from previous AoI scheduling instance, RRI$^{t-1}$\\%, $cap_{th}$ \\
\textbf{Output:} Returns true if the channel is congested as compared to the previous RRI, and false otherwise
\begin{algorithmic}[1]
\Procedure{$\textnormal{channel}_{\textnormal{cong}}$}{$s_{total}(1:N_{\rm sensing})$, $P_{th}^{t-1}$, RRI$^{t-1}$}
%\State{$\widehat{\textnormal{RRI}_v} = \frac{1}{\beta} \textnormal{ RRI}_{v}$} \Comment{New Potential RRI}
\State{$N_{\textnormal{RRI}}=\textnormal{floor}(\frac{N_{\rm sensing}}{\textnormal{RRI}^{t-1}}-1)$}

%\State{$T_2=\textnormal{RRI}^{t-1}$}
%\State {$S^{j}_{i}=\frac{1}{N_{\textnormal{RRI}}} [s^{j}_{1+i}, s^{j}_{1+i+\widehat{\textnormal{RRI}_v}},\cdots,s^{j}_{1+i+N_{\textnormal{RRI}}\cdot \widehat{\textnormal{RRI}_v}}]$}
\State {$S^{j}_{i}=\frac{1}{N_{\textnormal{RRI}}} \sum_{k=0}^{N_{RRI}} s^{j}_{1+i+k \cdot \textnormal{RRI}^{t-1}}$}

\State{$S_{\textnormal{total}}=[S^{j}_{0},S^{j}_{1},\dotsc, S^{j}_{\textnormal{RRI}^{t-1}}]$}
\If{$S^{j}_{i}<P_{th}^{t-1}$}
    \State {$S_{\textnormal{avail}} \leftarrow S^{j}_{i}$}
\EndIf
\If{$\frac{|S_{\textnormal{avail}}|}{|S_{\textnormal{total}}|}<0.2$}
    \State {return \textbf{true}}\Comment{The value of $P_{th}$ has not changed as compared to the previous slot selection, so the channel is not congested }

%\EndIf
\Else
\State {return \textbf{false}}\Comment{The value of $P_{th}$ needs to increase by $3$ dB as compared to the previous slot selection, so the channel is congested }
\EndIf
\EndProcedure
\end{algorithmic}
\end{algorithm}

\section{Simulation Overview and Results}
\label{Section- Simulation Description and Results}
% \begin{figure}
% \centering
% \includegraphics[scale=.4]{simulation_flowchartv2.jpeg}
% \caption{Flowchart showing layout of simulation}
% \centering
% \end{figure}
\vspace{-0.05in}

\par In this section we cover the simulation settings, performance metrics, and compare the proposed Ch-RRI SPS and AoI-RRI SPS to static RRI SPS.

% \begin{figure}[h]
% \centering
% \includegraphics[width=\columnwidth]{images/TTC_highway2.png}
% \caption{Highway scenario with time to collision showing "at risk" and "safe" vehicles}
% \label{system_model}
% \centering
% \end{figure}

\subsection{Simulation Setting}

We modified and enhanced a system-level simulator originally designed to model C-V2X Mode-4 \cite{LTEV2Vsim_paper} to model and compare AoI-RRI SPS, Ch-RRI SPS, and NR-V2X Mode-2 SPS with three different static RRIs. We use the 3GPP highway mobility models \cite{Tutorial_5G_NR_V2X} for our simulation results. %We use two models to validate our simulation models: the 3GPP highway mobility model, and urban mobility model. 
\par In the highway mobility model, vehicles move along a six lane highway, with three lanes dedicated to each direction. The highway models assume the velocity of vehicles moving in the positive direction and negative direction are drawn from truncated Gaussian distributions with means of $v_{avg}$ and $-v_{avg}$, respectively. The Gaussian distribution mean and variance values are assumed to be $19.44$ m/s ($70$ km/hr) and $3.0$ m/s, respectively. As each vehicle in each lane approaches the length of the highway, the warp used in the model places the vehicle at the opposite end of the highway. All vehicles remain in the same lane for the duration of the simulation. Each vehicle's initial location along the highway follows a Poisson distribution.  

\par For comprehensive analysis, we compare the performance of Ch-RRI SPS and AoI-RRI SPS against conventional NR-V2X Mode-2 SPS with three different static RRIs: $20$ ms RRI, $50$ ms RRI, and $100$ ms RRI. The vehicle densities in the simulation range from $20$ to $160$ veh/km, and initial positions and velocities do not change across AoI-RRI SPS, Ch-RRI SPS, and NR-V2X SPS. All simulations used a 10 MHz system bandwidth, $25$ second simulation time, and results were averaged over $10$ trials. Table \ref{TABLE_1} summarizes the default values of the key simulation parameters for AoI-RRI SPS, Ch-RRI SPS, and NR-V2X SPS with static RRIs.

\begin{table}
\caption{Simulation Parameters}
\label{TABLE_1}
\vspace{-0.1in}
\centering
%\small
\begin{tabular}{ |c|c|}
\hline
Parameter & Value\\
\hline
%Vehicle density & \{20,40,60,70,80,90,120,160\}\\&veh/km\\
Vehicle density & \{$20$-$160$\} veh/km\\\hline
Road Length/Number of lanes/Lane width & $2000$ m/ $6$ lanes/ $4$ m\\\hline
$\beta$ & $1.1$\\\hline
Reselection time & $0.5$ -$1.5$ s \\\hline
%Number of lanes & 6\\\hline
%Lane width & 4 m\\\hline
%\Vijay{$PDR$ distance bins} & \Vijay{\ 100 m}\\\hline
%Source-Dest Distance & 500 m\\\hline
%Safety beacon rate & 1.0- 10.0 Hz\\\hline
Simulation time & $25$ seconds\\\hline
Transmission Power & $23$ dBm\\\hline
Transmission and Sensing Range & $300$ meters\\\hline
%Sensing Range & 300 meters\\\hline
%Data Rate & 6 Mbps\\\hline
%Packet Length & 1000 bytes\\\hline
Distribution of vehicle speeds &  $N$($19.44$ m/s, $3$ m/s)\\\hline
%Deceleration $a$ & 4.6 m/$s^2$\\\hline
%$t_{react}$ & 1 second\\\hline
AoI-RRI/Ch-RRI RRI$_{max}$ & $100$ ms\\\hline
AoI-RRI/Ch-RRI RRI$_{min}$ & $20$ ms\\\hline
%\Vijay{Ch-RRI $cap_{th}$} & \Vijay{0.75}\\\hline
%\Vijay{RRI} & \Vijay{{20, 50, 100} ms} \\\hline
Packet Size & $190$ B\\\hline
MCS Index & $7$\\\hline
%Propagation Model & Winner+ B1 Model\\\hline
%$e_{th}$ \Vijay{What is $e_{th}$?} & 3 m \\\hline
Number of trials & 10\\\hline
Initial RSRP threshold, $P_{th}$ & $-90$ dBm\\\hline
%$thres_{min}$ & 5 s\\
%$thres_{max}$ & 15 s\\
\end{tabular}
\vspace{-0.2in}
\end{table}

 \begin{figure*}[!h]
 \vspace{-0.05in}
 \centering
 %\subfigure[\label{fig:IRT_comparison}]{
 %\epsfig{figure=images/IRT_comparison_total_v11.eps, width=2.3 in,  keepaspectratio}}
 \subfigure[\label{fig:TE_comparison}]{
 \epsfig{figure=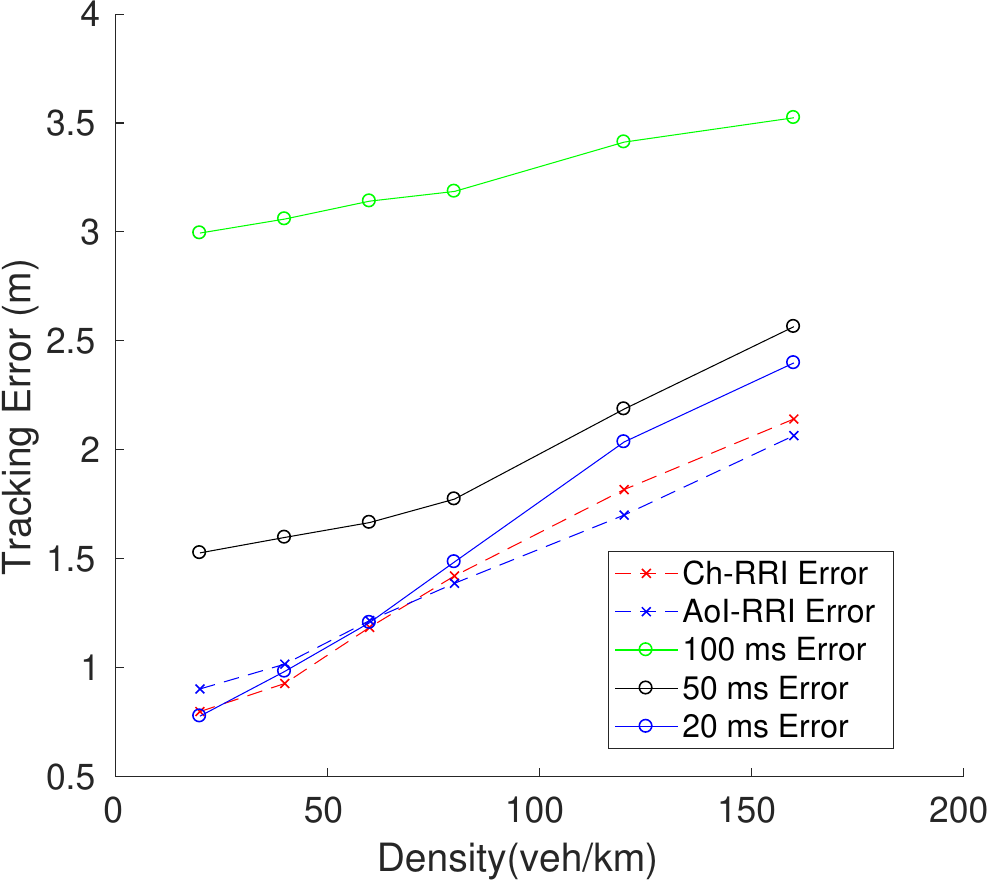, width=3.0 in,  keepaspectratio}}
  \subfigure[\label{fig:AOI_comparison}]{
 \epsfig{figure=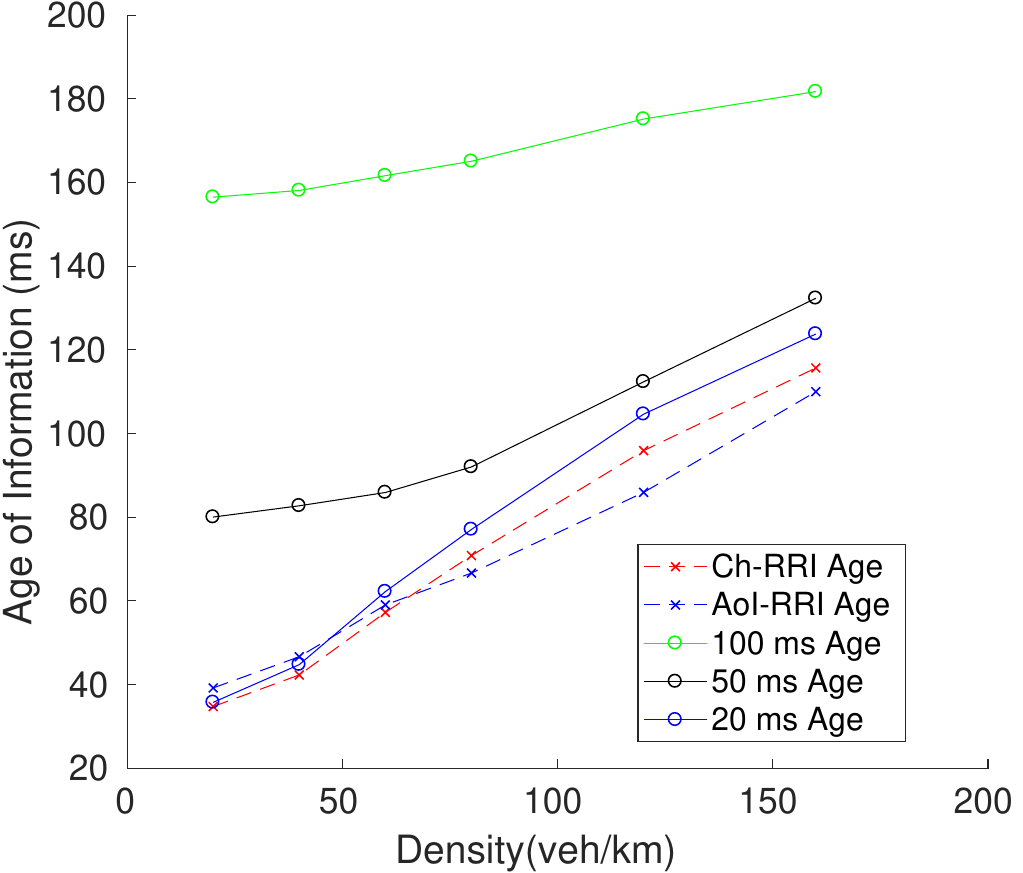, width=3.0 in,  keepaspectratio}}
 %\subfigure[\label{Tracking_error}]{
 %\epsfig{figure=images/HP_tracking_error_exp819_v2.eps,width=1.9 in,  keepaspectratio}}
 %\subfigure[\label{Collision_risk}]{
 %\epsfig{figure=images/collision_risk_exp819_v2.eps,width=1.9 in,  keepaspectratio}}
 \vspace{-0.1in}
 \caption{ (a) Average tracking error and (b) Average system age of information.}
 \vspace{-0.15in}
 \end{figure*}

\subsection{Performance Metrics}

\par This work uses the following three metrics to compare AoI-RRI SPS, Ch-RRI SPS, and static RRI SPS. %PDR and collision risk are the two major metrics used to evaluate the performance of Ch-RRI to fixed RRI SPS. 

\begin{itemize}
     \item \textbf{Tracking Error ($\mathbf{e_{track}}$)}- $e_{track}$ is calculated as the difference in the transmitting vehicle $u$'s actual location and estimated location from $u$'s last received transmission at receiver vehicle $v$. (see Section \ref{Subsection:Tracking Error}).
    %\item \textbf{Collision Risk} - The ratio of instances between a pair of transmitting ($v_t$) and receiving ($v_r$) vehicles where $v_t$'s tracking error $e_{tracking}$ exceeds the tracking error threshold $e_{tracking}$ while its time-to-collision ($TTC$) with $v_r$ is below the time to brake $t_{brake}$ to all other instances between $v_t$ and $v_r$. In this paper, the minimum time-to-collision and maximum tracking error allowed was 5.22 s and 2 m, respectively. 
    \item \textbf{Age of Information (AoI)} -  The difference in the reception time at receiver vehicle $v$ of vehicle $u$'s last location and the generation time of $u$'s last location (see Section \ref{Subsection:Age of Information}).
    
    %At each millisecond, the simulation monitors the tracking error and time-to-collision for every transmitting vehicle $v$ and receiving vehicle $u$, using Eq.  \ref{Collision_risk_eq_2} to decide whether a \textbf{collision risky instance} ever occurs between vehicle $v$ and $u$. The simulation then computes \textbf{collision risk} as the \textbf{ratio} of collision risky instances to all other instances between vehicle $v$ and $u$.
    
    \item  \textbf{Packet Delivery Ratio (PDR)} - The likelihood that all neighbors inside the transmission range of a vehicle successfully receive the transmitted packet. Formally, the PDR is computed as PDR$_u=\frac{PR_i}{PD_i}$. $PD_i$ is the number of packets sent by vehicle $i$ and $PR_i$ is the number of packets received by neighboring vehicles originally transmitted by vehicle $i$. 
    %The PDR is computed every second of the simulation, and classifies packets into varied distance bins around each 
\end{itemize}

\subsection{Experimental Results}
%Figs. 2-4 show the PDR results of the 
 \begin{figure*}[h]
 %\vspace{-0.05in}
 \centering
 \subfigure[$20$ veh/km \label{fig:Distributions-20 veh}]{
 \epsfig{figure=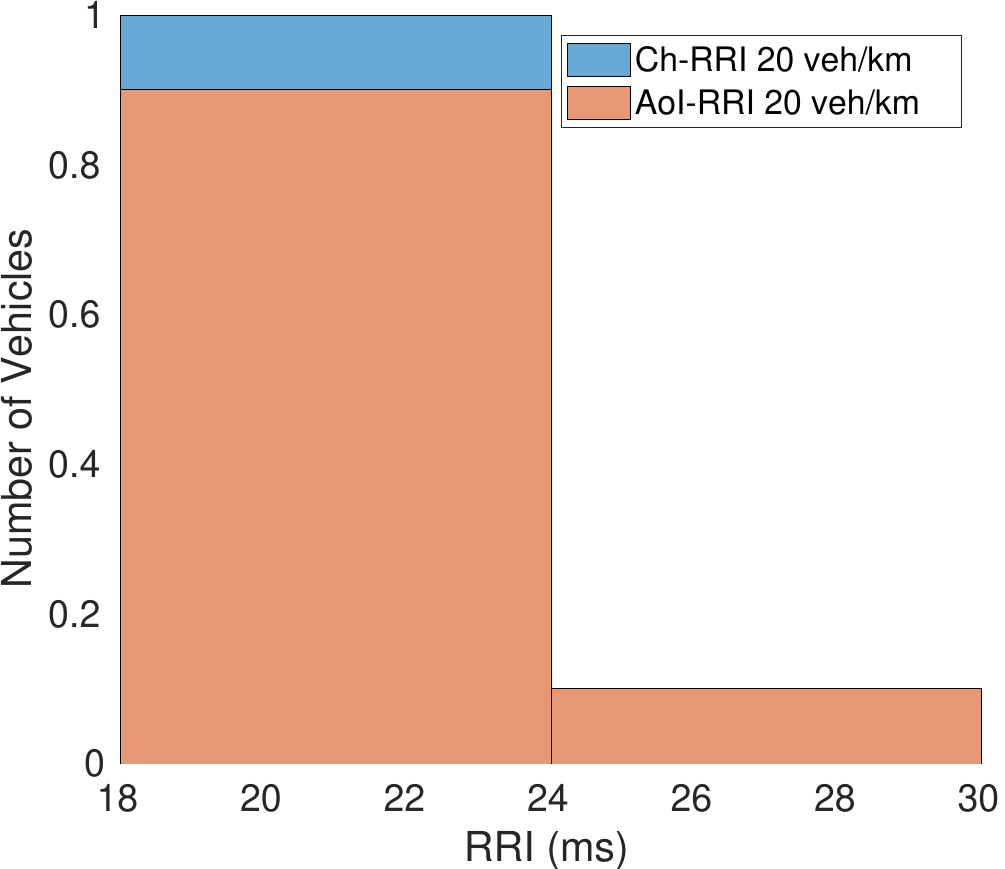, width=1.7 in,  keepaspectratio}}
 \subfigure[$60$ veh/km \label{fig:Distributions-60 veh}]{
 \epsfig{figure=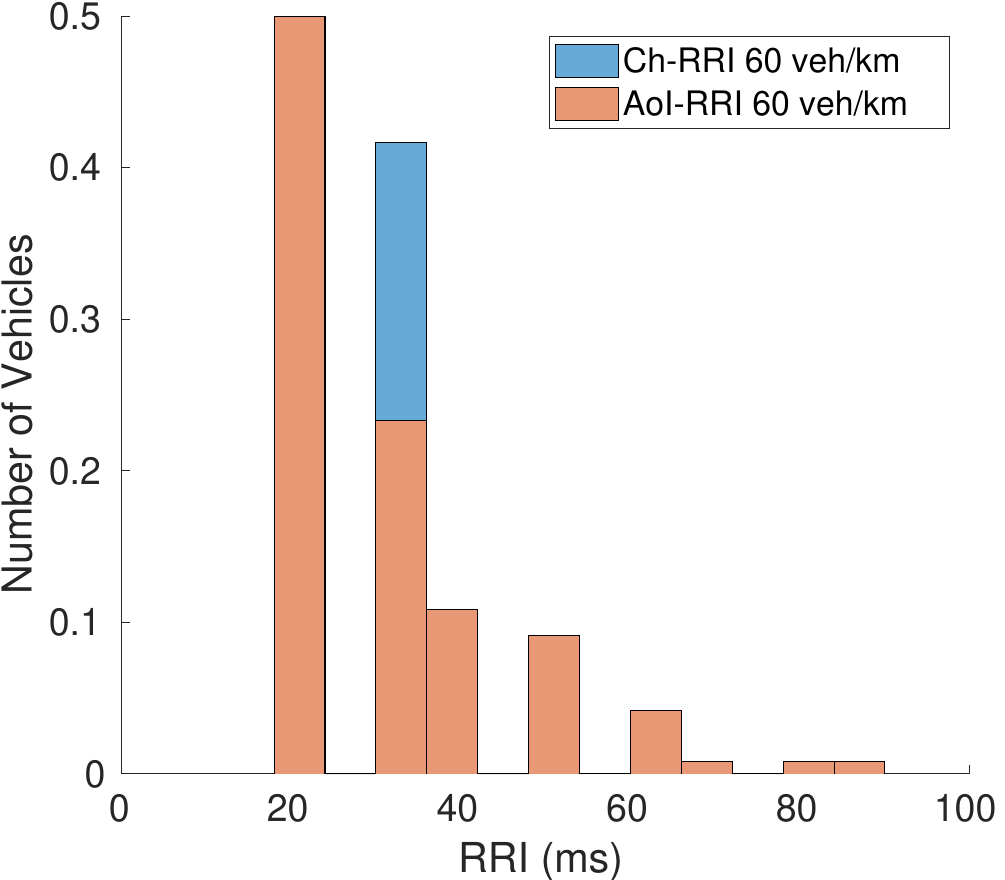, width=1.7 in,  keepaspectratio}}
 \subfigure[$120$ veh/km \label{fig:Distributions-120 veh}]{
\epsfig{figure=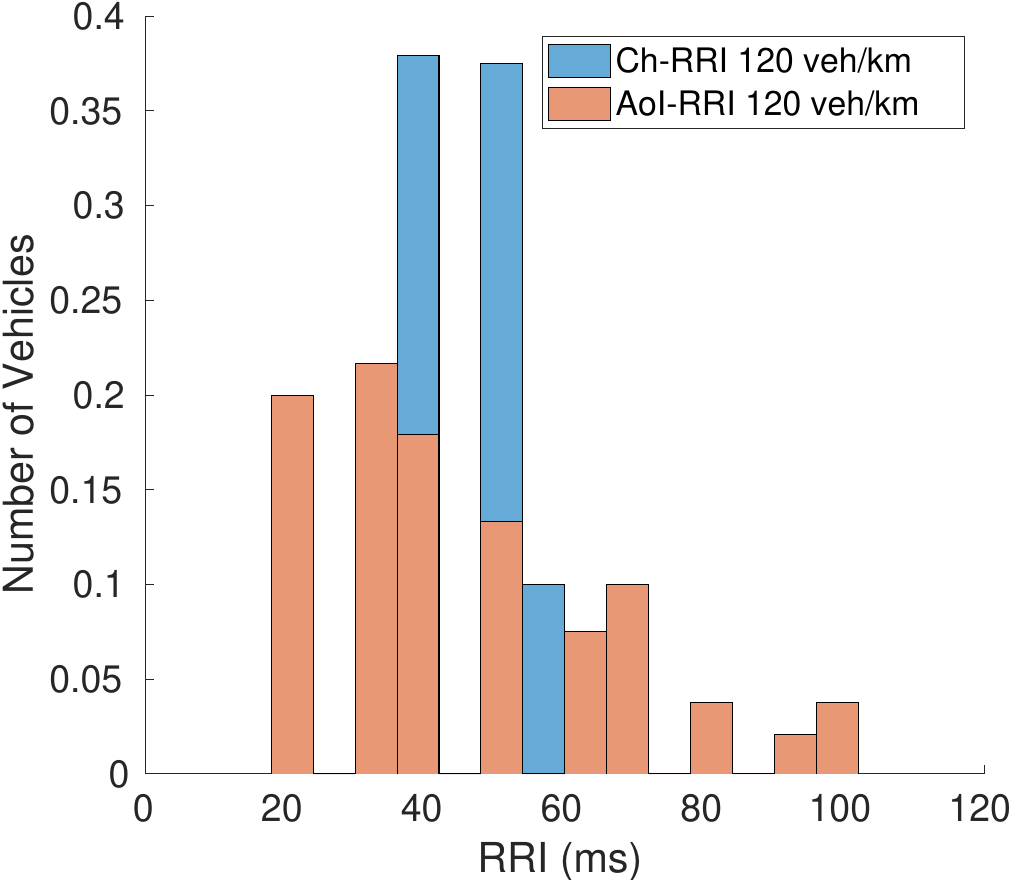,width=1.7 in,  keepaspectratio}}
  \subfigure[$160$ veh/km \label{fig:Distributions-160 veh}]{
 \epsfig{figure=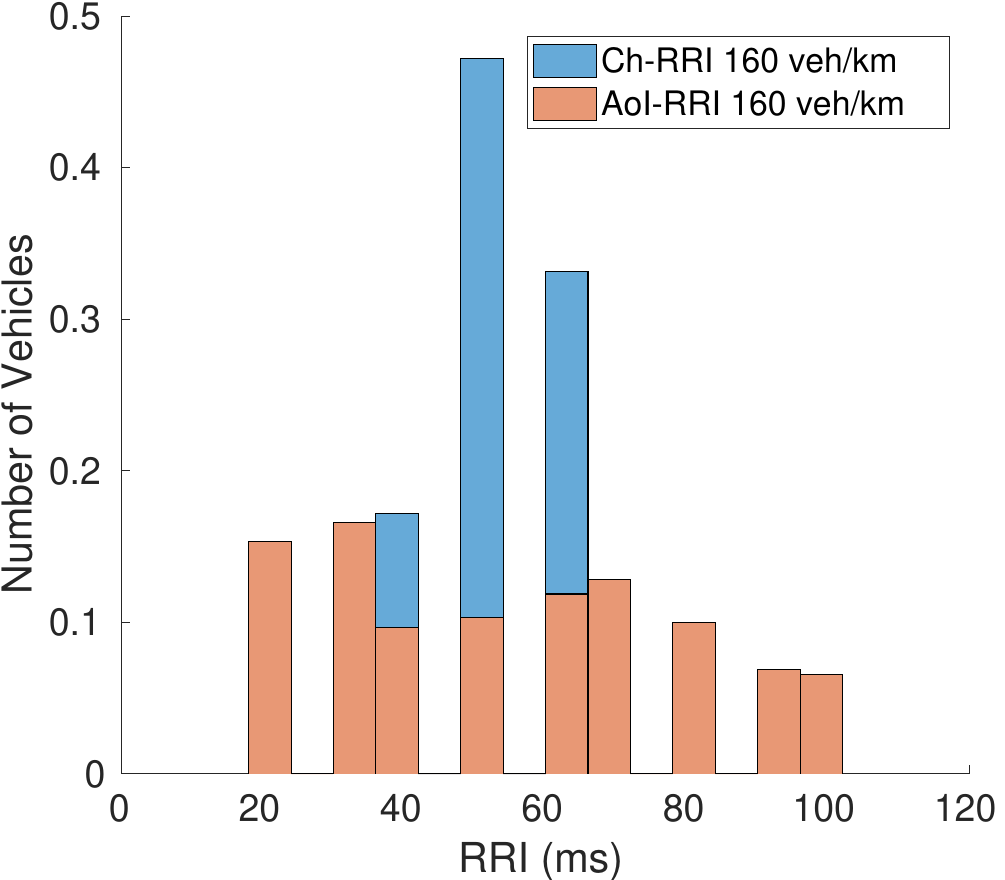, width=1.7 in,  keepaspectratio}}
 %\subfigure[\label{Tracking_error}]{
 %\epsfig{figure=images/HP_tracking_error_exp819_v2.eps,width=1.9 in,  keepaspectratio}}
 %\subfigure[\label{Collision_risk}]{
 %\epsfig{figure=images/collision_risk_exp819_v2.eps,width=1.9 in,  keepaspectratio}}
 %\vspace{-0.1in}
 \caption{ RRI distribution during the last 5 seconds of simulations across vehicle densities for Ch-RRI and AoI-RRI.}
 %\vspace{-0.15in}
 \end{figure*}
  \begin{figure*}[h]
 %\vspace{-0.14in}
 \subfigure[$20$ veh/km \label{fig:20_veh_AoI_time}]{
 \epsfig{figure=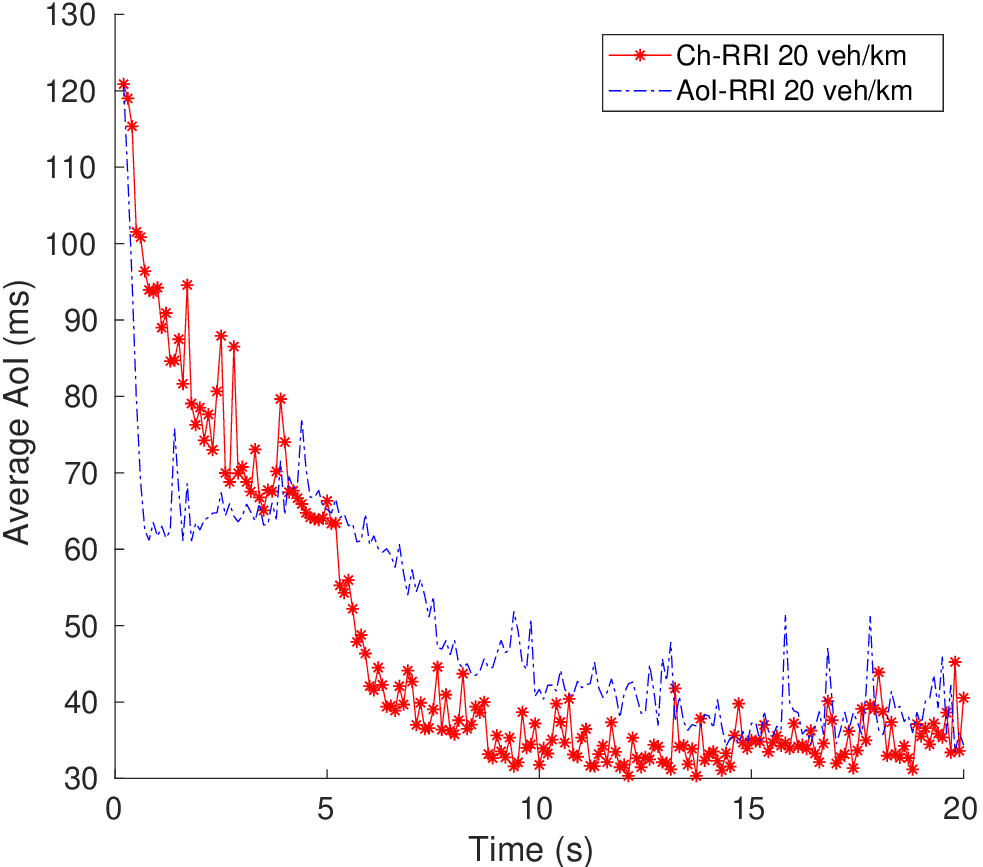,width=1.7 in,  keepaspectratio}}
  \subfigure[$60$ veh/km \label{fig:60_veh_AoI_time}]{
 \epsfig{figure=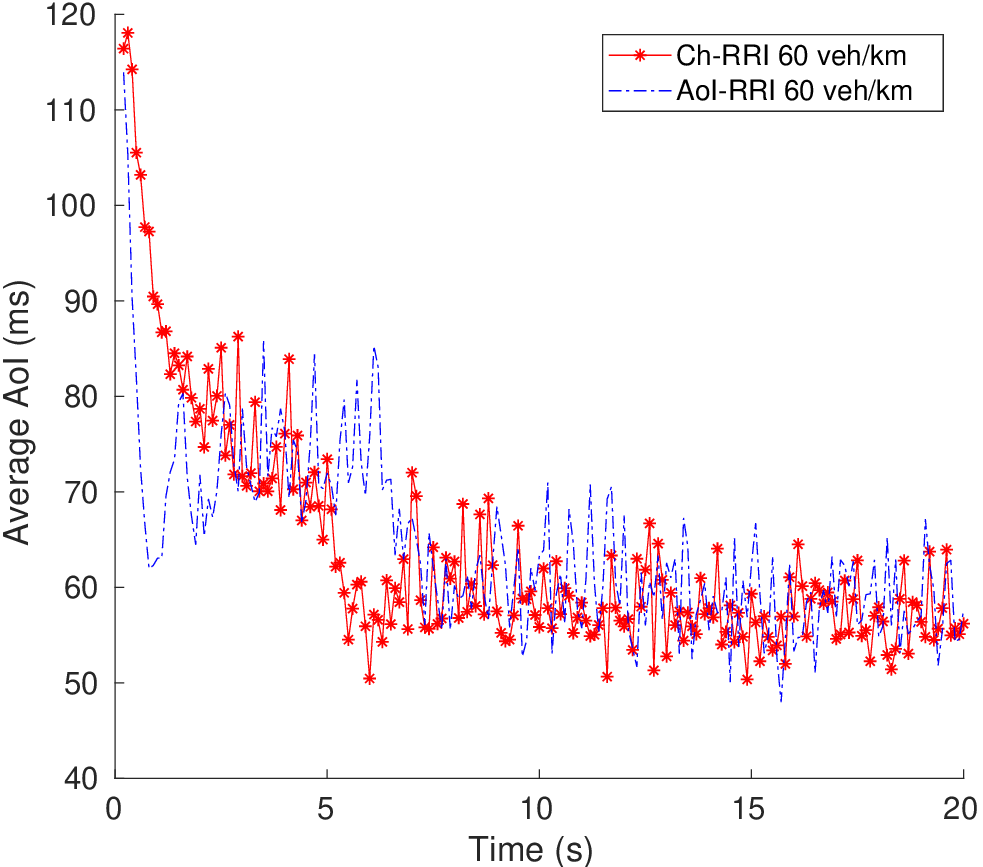,width=1.7 in,  keepaspectratio}}
   \subfigure[$120$ veh/km \label{fig:120_veh_AoI_time}]{
  \epsfig{figure=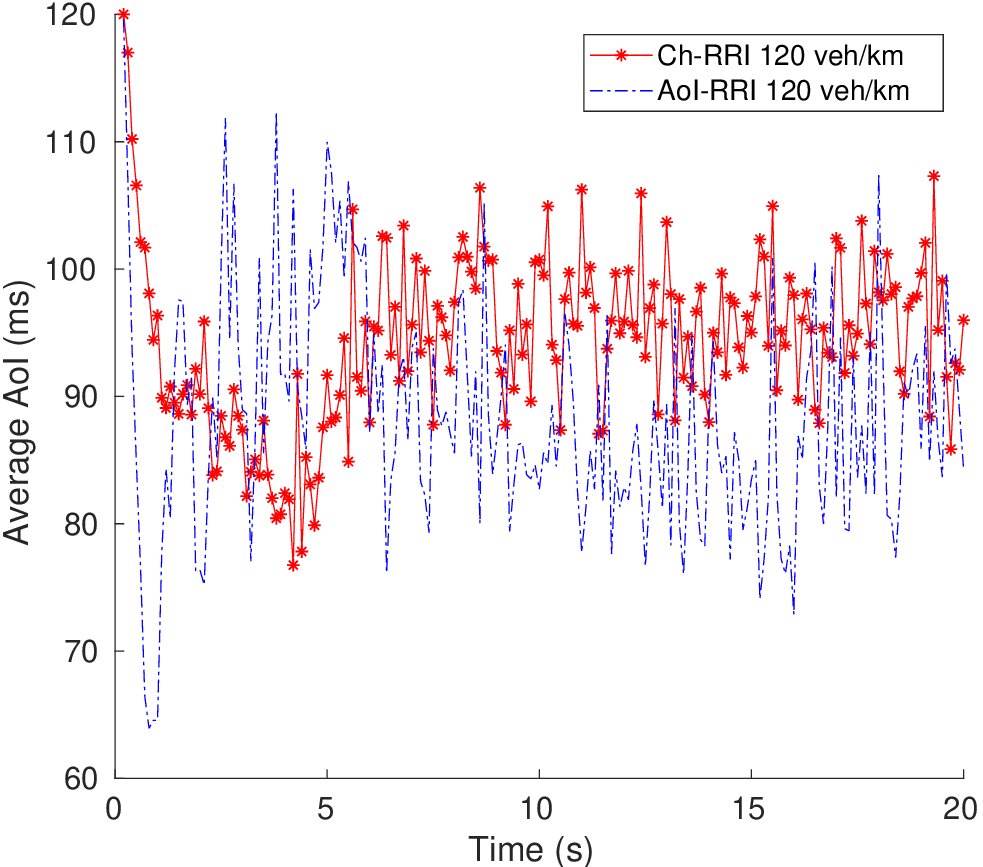,width=1.7 in,  keepaspectratio}}
 \subfigure[$160$ veh/km \label{fig:160_veh_AoI_time}]{
 \epsfig{figure=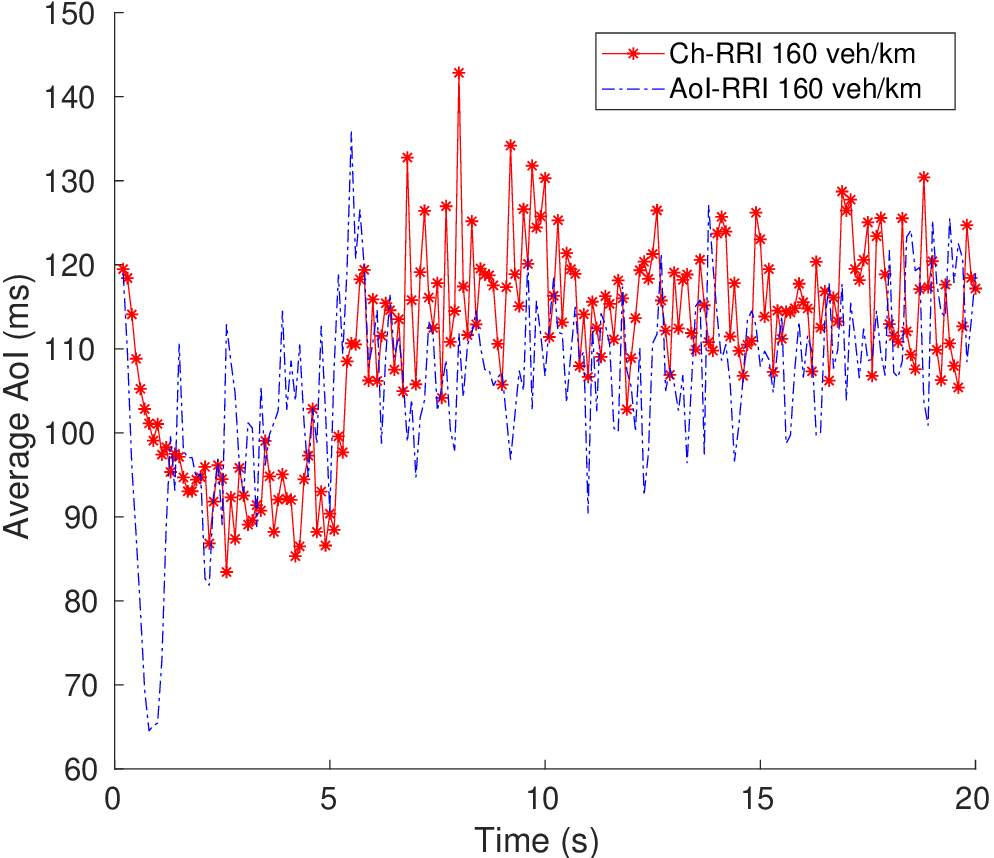,width=1.7 in,  keepaspectratio}}
 % \vspace{-0.09in}
 \caption{Average AoI vs time across vehicle densities for Ch-RRI and AoI-RRI.} \label{fig:AoI_time_results}
 %\vspace{-0.10in}
 \end{figure*}

 \begin{figure*}[h]
 %\vspace{-0.14in}
 \subfigure[$20$ veh/km \label{fig:20_veh_PDR}]{
 \epsfig{figure=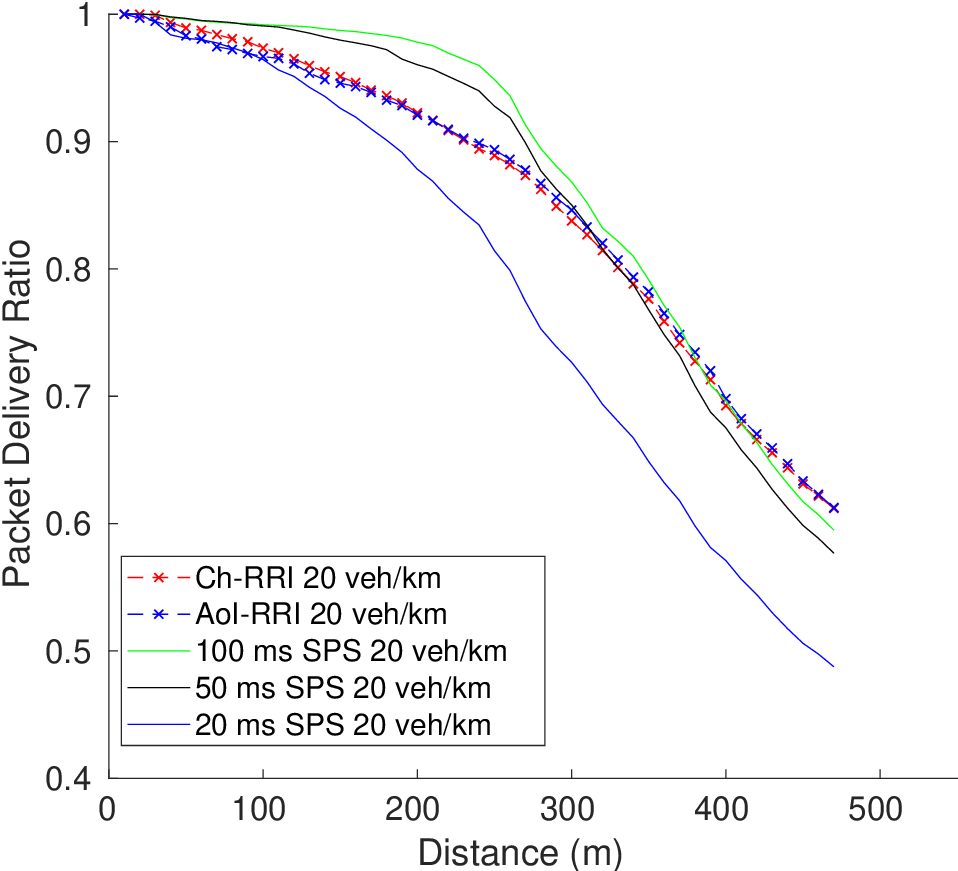,width=1.7 in,  keepaspectratio}}
  \subfigure[$60$ veh/km \label{fig:60_veh_PDR}]{
 \epsfig{figure=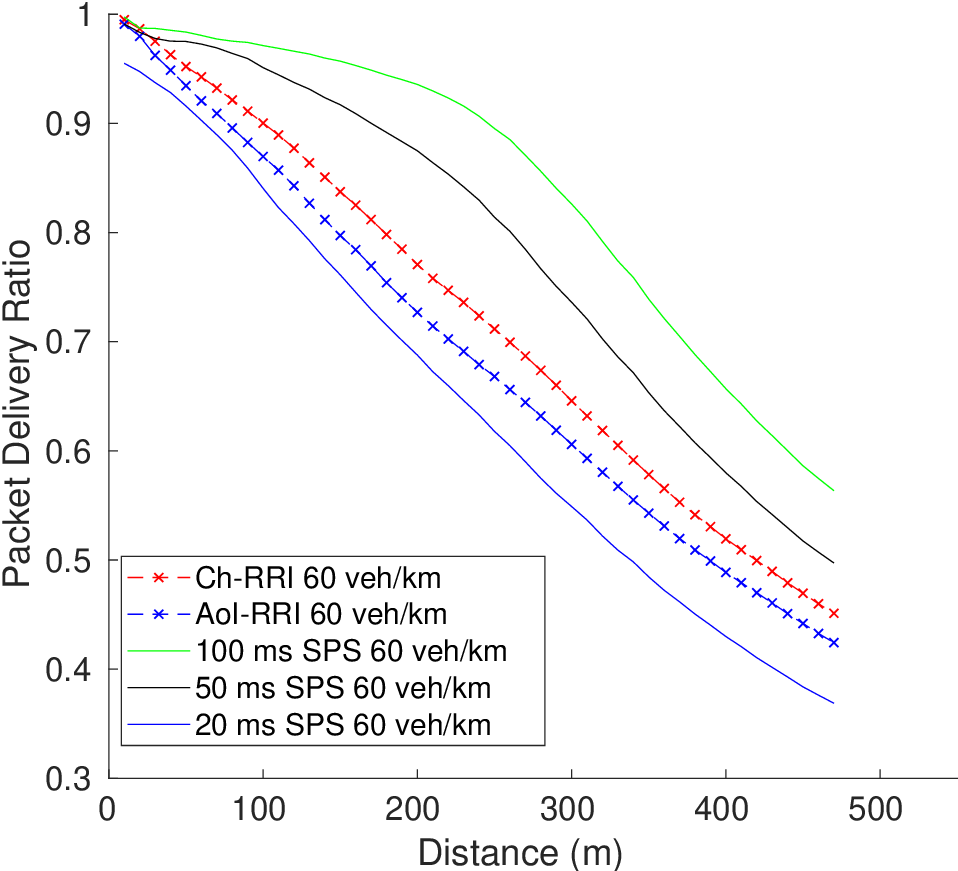,width=1.7 in,  keepaspectratio}}
  \subfigure[$120$ veh/km \label{fig:120_veh_AoI_time}]{
  \epsfig{figure=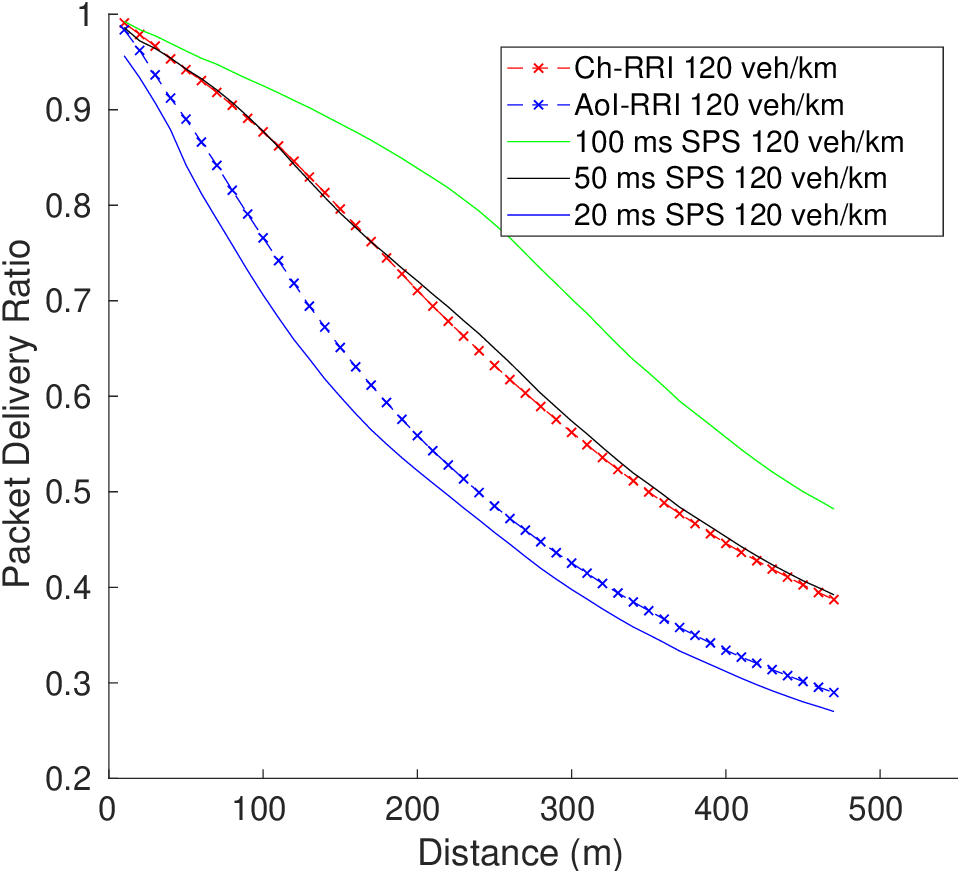,width=1.7 in,  keepaspectratio}}
 \subfigure[$160$ veh/km \label{fig:160_veh_PDR}]{
 \epsfig{figure=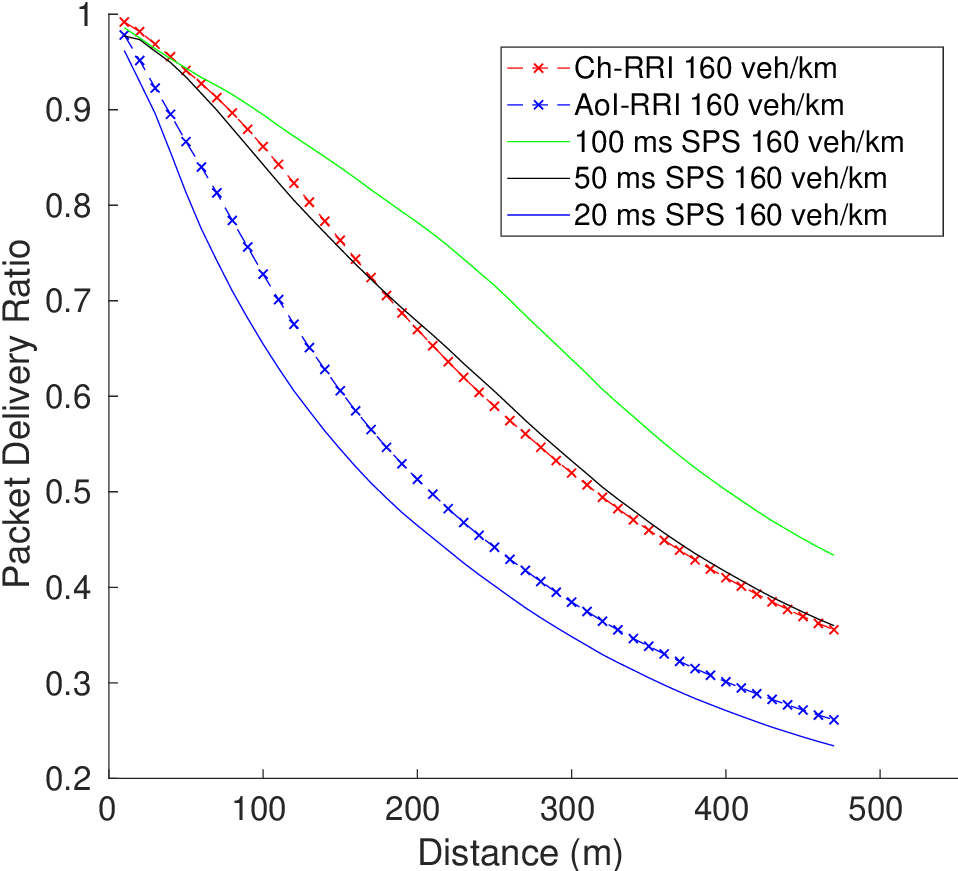,width=1.7 in,  keepaspectratio}}
  \vspace{-0.09in}
 \caption{ PDR across vehicle densities.} \label{fig:PDR_results}
 \vspace{-0.10in}
 \end{figure*} 
We compare AoI-RRI, Ch-RRI, and static RRI SPS using the aforementioned tracking error, AoI, and PDR performance metrics. In addition to these metrics, we also look at the final RRI distributions and average AoI over time to better understand the operation of AoI-RRI and Ch-RRI. %We then detail how AoI-RRI and Ch-RRI select each vehicle's RRI over varying vehicle densities and time.

%Fig. \ref{fig:RRI_time} shows how AoI-RRI chooses, on average, the RRI across all vehicles for each considered vehicle density scenario (i.e., $40$, $80$, $120$ and $160$) under varying simulation time.  We observe over 10 trials that the average chosen RRI converges over simulation time as vehicles enter the simulation for each considered vehicle density. After $10$ seconds, the RRI chosen for each vehicle setting converges to a distribution of RRI values. The chosen RRI in case of $40$ and $80$ vehicle density (i.e., sparse) is almost four to five times than that of $120$ or $160$ (i.e., dense) vehicle setting, thanks to the ability of AoI-RRI to adjust RRI in real-time and adapt to the varying C-V2X environment. Vehicles also take a longer time in dense vehicle settings ($5$ and $6$ seconds for 120 and 160 vehicles respectively) to converge to a chosen RRI value, likely because it takes longer for all vehicles to enter the simulation.   
\textbf{Average Tracking Error and System \textit{AoI}.}  Figs. \ref{fig:TE_comparison} and \ref{fig:AOI_comparison} compare the system tracking error and AoI, respectively, of AoI-RRI SPS and Ch-RRI SPS to $20$ ms, $50$ ms, and $100$ ms static RRI SPS. Fig. \ref{fig:TE_comparison} demonstrates that tracking error worsens with increasing density, with the $20$ ms RRI SPS tracking error giving the best static RRI SPS performance. Both tracking error and AoI increase drastically at higher densities (larger than 100 veh/km). Since there are insufficient slot resources to support a $20$ ms RRI without two vehicles in the same vicinity choosing the same slot resource, there are increased packet losses, leading to worse performance at high densities. The tracking error for AoI-RRI is smaller than the tracking error of Ch-RRI and all three static RRIs across large vehicle densities. AoI-RRI outperforms $20$ ms RRI NR-V2X (the best performing fixed RRI) in terms of AoI by almost $19\%$ and $16\%$ respectively at $120$ and $160$ veh/km. This indicates that AoI-RRI finds for each vehicle a RRI that minimizes packet losses and enables timely cooperative awareness in the network. The AoI improvements found with AoI-RRI extend to the tracking error. AoI-RRI was also found to have an advantage over Ch-RRI in high density scenarios as well outperforming Ch-RRI  at 120 and 160 veh/km by $9.16\%$ and $10.81\%$, respectively.  
The tracking error results show that lower average RRIs aid vehicles in high and low density situations. Even in high density scenarios, ($160$ vehicles/km), both AoI-RRI  and Ch-RRI outperform the $20$ and $50$ ms RRI. 

\textbf{RRI Distribution.} Figs. \ref{fig:Distributions-20 veh}-\ref{fig:Distributions-160 veh} show the average RRI distributions for AoI-RRI and Ch-RRI for selected densities over the last five seconds of the simulation\footnote{We use the last five seconds because by then, every vehicle has had sufficient time to choose an acceptable RRI and we can analyze steady state behavior.}. Across densities, AoI-RRI  shows a larger RRI variance as compared to the Ch-RRI. This large variance in the RRI distribution for the AoI-RRI algorithm can be attributed to varying traffic densities and noisy estimates of the measured AoI which leads vehicles to choose different RRIs depending on the observed AoI. Although there is a large variance in the RRI distribution of the AoI-RRI, the AoI-RRI tracking error  performance approaches that of Ch-RRI and $20$ ms SPS for low densities, and outperforms all methods in high density scenarios.
In Ch-RRI, the RRI distribution is comparatively narrow, and the average chosen RRI is similarly small for low densities ($20$ and $40$ veh/km). This means that there are enough channel resources to support small RRIs for both Ch-RRI and AoI-RRI. In Fig. \ref{fig:Distributions-160 veh}, we see that while the AoI-RRI RRI distribution skews towards RRI$_{max}$, many vehicles use lower RRIs, which contributes to the improved AoI performance.
\par \textbf{Average \textit{AoI} vs Time.} 
Figs. \ref{fig:20_veh_AoI_time}-\ref{fig:160_veh_AoI_time} present the average AoI over simulation time of AoI-RRI SPS and Ch-RRI SPS. Both AoI-RRI and Ch-RRI algorithms start with an initial RRI of 50 ms, and only allow for adaptive RRIs to be chosen after 5 sec. This gives enough time for each algorithm to (i) allow all vehicles to transmit at least once and (ii) allow each algorithm to gather channel and AoI measurements. Notice that both AoI-RRI and Ch-RRI start with an initial value of 120 ms. This is because both algorithms start with observing the channel for 100 ms (following the NR-V2X Mode-2 SPS standard) before selecting an initial slot. Once both algorithms start transmitting, the AoI drops to 70 ms, which is expected for an RRI of 50 ms. In the low density cases (20 and 40 veh/km), both algorithms drop to average AoI of 40-50 ms. Note that Ch-RRI slightly outperforms AoI-RRI in lower densities. This is because Ch-RRI detects an empty channel and automatically transmits at the lowest possible RRI. Ch-RRI is also able to converge to a steady state value faster than AoI-RRI, likely because of the noisy nature of the AoI estimates. When the density increases, AoI-RRI is able to perform better, and is able to take actions that lead to a lower average AoI.

\textbf{Packet Delivery Ratio.} Fig. \ref{fig:PDR_results} compares the AoI-RRI, Ch-RRI, and NR-V2X Mode-2 SPS PDRs across vehicle densities. Among the fixed RRI results, the $100$ ms RRI performs the best, but correspondingly yield a large tracking error and AoI. Similarly, $100$ ms RRI NR-V2X gives the worst PDR performance, but gave the best tracking error and AoI performance, meaning there is a tradeoff between PDR and RRI, that affects AoI performance. Both AoI-RRI and Ch-RRI attempt to find the best RRI that optimizes the tracking error and AoI. 
\\
Notice that at low densities (20 and 60 veh/km), the PDR of AoI-RRI and Ch-RRI are similar, as are the average AoI and tracking error. Both AoI-RRI and Ch-RRI choose similar RRIs and are able to achieve the best tracking error performance. As the densities increase the RRI distribution of AoI-RRI tends towards lower RRIs, and the PDR performance is worse than that of Ch-RRI, which chooses a larger average RRI. This indicates that the AoI-RRI finds that choosing a slightly lower RRI distribution can minimize the AoI, as the average age of AoI-RRI  is lower than Ch-RRI. However, AoI-RRI  does not select a $20$ ms RRI for all vehicular since the AoI results show there are diminishing returns for selecting a low RRI, and eventually the increased congestion and packet collisions will deteriorate AoI and tracking error.

\textbf{Discussion.} Unlike conventional SPS and Ch-RRI, the proposed AoI aware SPS is able to successfully learn the system AoI and adapt each vehicle's RRI across time-varying NR-V2X scenarios. 
As a result of choosing an optimal RRI, it is shown that the average tracking error of every vehicle pair is also reduced. This significantly improves the cooperative awareness of vehicular networks.
\section{Conclusion} \label{Section-Conclusion}
%\Avik{highlight adaptive RRI, optimal ultilization of resources, improve safety}

In this work, we proposed two adaptive RRI algorithms to power SPS for improved cooperative awareness performance of decentralized V2X networks. Ch-RRI and AoI-RRI choose RRIs based on the availability of channel resources and measured AoI, respectively, and then use  SPS for the selection of suitable BSM transmission opportunities at those chosen RRIs. Our extensive experiments based on NR-V2X Mode-2 standard demonstrated that SPS powered by either algorithm significantly outperforms conventional SPS in terms of the cooperative awareness performance in all considered NR-V2X scenarios. In the future, we will explore using nonlinear age functions and designing reinforcement learning (RL) based scheduling protocols that can learn vehicle priorities and other contextual factors over time.

%\begin{equation}
%    \sum_{c=1}^{k} c\left(1-p_{s}\right)^{c-1}=\frac{1-\left(1+k p_{s}\right) \cdot\left(1-p_{s}\right)^{k}}{p_{s}^{2}}
%\end{equation}

\bibliographystyle{IEEEtran}
%\bibliography{bibliography}
%
\bibliography{main_short}

\begin{thebibliography}{10}
\providecommand{\url}[1]{#1}
\csname url@samestyle\endcsname
\providecommand{\newblock}{\relax}
\providecommand{\bibinfo}[2]{#2}
\providecommand{\BIBentrySTDinterwordspacing}{\spaceskip=0pt\relax}
\providecommand{\BIBentryALTinterwordstretchfactor}{4}
\providecommand{\BIBentryALTinterwordspacing}{\spaceskip=\fontdimen2\font plus
\BIBentryALTinterwordstretchfactor\fontdimen3\font minus
  \fontdimen4\font\relax}
\providecommand{\BIBforeignlanguage}[2]{{%
\expandafter\ifx\csname l@#1\endcsname\relax
\typeout{** WARNING: IEEEtran.bst: No hyphenation pattern has been}%
\typeout{** loaded for the language `#1'. Using the pattern for}%
\typeout{** the default language instead.}%
\else
\language=\csname l@#1\endcsname
\fi
#2}}
\providecommand{\BIBdecl}{\relax}
\BIBdecl

\bibitem{IFIP_networking_paper}
A.~Dayal, V.~K. Shah, B.~Choudhury, V.~Marojevic, C.~Dietrich, and J.~H. Reed,
  ``{Adaptive Semi-Persistent Scheduling for Enhanced On-road Safety in
  Decentralized V2X Networks},'' in \emph{Proc., IFIP Netw. Conf.}, 2021, pp.
  1--9.

\bibitem{dayal2021practical}
A.~Dayal, ``Practical algorithms and analysis for next-generation decentralized
  vehicular networks,'' Ph.D. dissertation, Virginia Tech, 2021.

\bibitem{mmwave_IOV_IOTJ_ghafoor}
K.~Zrar~Ghafoor, L.~Kong, S.~Zeadally, A.~S. Sadiq, G.~Epiphaniou,
  M.~Hammoudeh, A.~K. Bashir, and S.~Mumtaz, ``{Millimeter-Wave Communication
  for Internet of Vehicles: Status, Challenges, and Perspectives},'' \emph{IEEE
  Internet Things J.}, vol.~7, no.~9, pp. 8525--8546, 2020.

\bibitem{Collective_perception_Thandavarayan_2019}
G.~Thandavarayan, M.~Sepulcre, and J.~Gozalvez, ``{Analysis of Message
  Generation Rules for Collective Perception in Connected and Automated
  Driving},'' in \emph{2019 IEEE Intelligent Vehicles Symposium (IV)}, 2019,
  pp. 134--139.

\bibitem{V2V_safety_coop_perception}
S.-W. Kim, W.~Liu, M.~H. Ang, E.~Frazzoli, and D.~Rus, ``{The impact of
  cooperative perception on decision making and planning of autonomous
  vehicles},'' \emph{IEEE Intell. Transp. Syst. Mag.}, vol.~7, no.~3, pp.
  39--50, 2015.

\bibitem{Naik_survey_5G}
G.~{Naik}, B.~{Choudhury}, and J.~{Park}, ``{IEEE 802.11bd and 5G NR V2X:
  Evolution of Radio Access Technologies for V2X Communications},'' \emph{IEEE
  Access}, vol.~7, pp. 70\,169--70\,184, 2019.

\bibitem{molina2020comparison}
R.~Molina-Masegosa, J.~Gozalvez, and M.~Sepulcre, ``{Comparison of IEEE 802.11
  p and LTE-V2X: An evaluation with periodic and aperiodic messages of constant
  and variable size},'' \emph{IEEE Access}, vol.~8, pp. 121\,526--121\,548,
  2020.

\bibitem{magazine_paper}
R.~{Molina-Masegosa} and J.~{Gozalvez}, ``{LTE-V for Sidelink 5G V2X Vehicular
  Communications: A New 5G Technology for Short-Range Vehicle-to-Everything
  Communications},'' \emph{IEEE Veh. Technol. Mag.}, vol.~12, no.~4, pp.
  30--39, 2017.

\bibitem{Tutorial_5G_NR_V2X}
M.~H.~C. {Garcia}, A.~{Molina-Galan}, M.~{Boban}, J.~{Gozalvez},
  B.~{Coll-Perales}, T.~{Şahin}, and A.~{Kousaridas}, ``{A Tutorial on 5G NR
  V2X Communications},'' \emph{IEEE Commun. Surv. Tutor.}, pp. 1--1, 2021.

\bibitem{ETSI_EN_302}
``{ETSI EN 302 637-2 V1.3.2 (2014-11) Intelligent Transport Systems (ITS);
  Vehicular Communications; Basic Set of Applications; Part 2: Specification of
  Cooperative Awareness Basic Service.}''

\bibitem{milanes2013cooperative}
V.~Milan{\'e}s, S.~E. Shladover, J.~Spring, C.~Nowakowski, H.~Kawazoe, and
  M.~Nakamura, ``{Cooperative adaptive cruise control in real traffic
  situations},'' \emph{IEEE Trans. Intell. Transp. Syst.}, vol.~15, no.~1, pp.
  296--305, 2013.

\bibitem{Huang_IEEENetwork}
C.~{Huang}, Y.~P. {Fallah}, R.~{Sengupta}, and H.~{Krishnan}, ``{Adaptive
  intervehicle communication control for cooperative safety systems},''
  \emph{IEEE Netw.}, vol.~24, no.~1, pp. 6--13, 2010.

\bibitem{rasouli2019autonomous}
A.~Rasouli and J.~K. Tsotsos, ``{Autonomous vehicles that interact with
  pedestrians: A survey of theory and practice},'' \emph{IEEE Trans. Intell.
  Transp. Syst.}, vol.~21, no.~3, pp. 900--918, 2019.

\bibitem{Biplav_VTC_2020}
B.~Choudhury, V.~K. Shah, A.~Dayal, and J.~H. Reed, ``{Experimental Analysis of
  Safety Application Reliability in V2V Networks},'' in \emph{Proc., IEEE Veh.
  Technol. Conf. (VTC)}, 2020, pp. 1--5.

\bibitem{IOTJ_2016_V2X_Chen}
S.~{Chen}, J.~{Hu}, Y.~{Shi}, and L.~{Zhao}, ``{LTE-V: A TD-LTE-Based V2X
  Solution for Future Vehicular Network},'' \emph{IEEE Internet Things J.},
  vol.~3, no.~6, pp. 997--1005, 2016.

\bibitem{Nabil_SPS}
A.~{Nabil}, K.~{Kaur}, C.~{Dietrich}, and V.~{Marojevic}, ``{Performance
  Analysis of Sensing-Based Semi-Persistent Scheduling in C-V2X Networks},'' in
  \emph{Proc., IEEE Veh. Technol. Conf. (VTC)}, 2018, pp. 1--5.

\bibitem{Bazzi_CV2X}
A.~{Bazzi}, G.~{Cecchini}, A.~{Zanella}, and B.~M. {Masini}, ``{Study of the
  Impact of PHY and MAC Parameters in 3GPP C-V2V Mode 4},'' \emph{IEEE Access},
  vol.~6, pp. 71\,685--71\,698, 2018.

\bibitem{IOTJ_SPS}
X.~{He}, J.~{Lv}, J.~{Zhao}, X.~{Hou}, and T.~{Luo}, ``{Design and Analysis of
  a Short Term Sensing Based Resource Selection Scheme for C-V2X Networks},''
  \emph{IEEE Internet Things J.}, pp. 1--1, 2020.

\bibitem{bazzi2019congestion}
A.~Bazzi, ``{Congestion control mechanisms in IEEE 802.11 p and sidelink
  C-V2X},'' in \emph{Proc., IEEE Asilomar}, 2019, pp. 1125--1130.

\bibitem{Peng_TWC_AOI_optimized_mac}
F.~{Peng}, Z.~{Jiang}, S.~{Zhang}, and S.~{Xu}, ``{Age of Information Optimized
  MAC in V2X Sidelink via Piggyback-Based Collaboration},'' \emph{IEEE Trans.
  on Wireless Commun.}, vol.~20, no.~1, pp. 607--622, 2021.

\bibitem{Bexmenov_SPS_Aperiodic}
M.~Bezmenov, Z.~Utkovski, K.~Sambale, and S.~Stanczak, ``{Semi-Persistent
  Scheduling with Single Shot Transmissions for Aperiodic Traffic},'' in
  \emph{Proc., IEEE Veh. Technol. Conf. (VTC)}, 2021, pp. 1--7.

\bibitem{Analytical_CV2X}
M.~{Gonzalez-Martín}, M.~{Sepulcre}, R.~{Molina-Masegosa}, and J.~{Gozalvez},
  ``{Analytical Models of the Performance of C-V2X Mode 4 Vehicular
  Communications},'' \emph{IEEE Trans. on Veh. Technology}, vol.~68, no.~2, pp.
  1155--1166, 2019.

\bibitem{campolo20195g}
C.~Campolo, A.~Molinaro, F.~Romeo, A.~Bazzi, and A.~O. Berthet, ``{5G NR V2X:
  On the impact of a flexible numerology on the autonomous sidelink mode},'' in
  \emph{Proc. of IEEE 2nd 5G World Forum (5GWF)}.\hskip 1em plus 0.5em minus
  0.4em\relax IEEE, 2019, pp. 102--107.

\bibitem{Zoraze_NRV2X_Mode2_overview}
Z.~Ali, S.~Lagén, L.~Giupponi, and R.~Rouil, ``{3GPP NR V2X Mode 2: Overview,
  Models and System-Level Evaluation},'' \emph{IEEE Access}, vol.~9, pp.
  89\,554--89\,579, 2021.

\bibitem{Bartoletti_Access_2021}
S.~{Bartoletti}, B.~{Masini}, C.~{Campolo}, V.~{Martinez}, S.~{Ioannis}, and
  A.~{Bazzi}, ``Impact of the generation interval on the performance of
  sidelink c-v2x autonomous mode,'' \emph{IEEE Access}, vol.~9, pp.
  35\,121--35\,135, 2021.

\bibitem{Todisco_Access_2021}
V.~{Todisco}, S.~{Bartoletti}, C.~{Campolo}, A.~{Molinaro}, A.~{Berthet}, and
  A.~{Bazzi}, ``Performance analysis of sidelink 5g-v2x mode 2 through an
  open-source simulator,'' \emph{IEEE Access}, vol.~9, pp. 145\,648--145\,661,
  2021.

\bibitem{Lee_TENCON_2020}
T.-H. Lee and C.-F. Lin, ``{Reducing Collision Probability in Sensing-Based SPS
  Algorithm for V2X Sidelink Communications},'' in \emph{Proc., IEEE Region 10
  Conf. (TENCON)}, 2020, pp. 303--308.

\bibitem{kaul2012real}
S.~Kaul, R.~Yates, and M.~Gruteser, ``{Real-time status: How often should one
  update?}'' in \emph{Proc., IEEE INFOCOM}.\hskip 1em plus 0.5em minus
  0.4em\relax IEEE, 2012, pp. 2731--2735.

\bibitem{baiocchi2020age}
A.~Baiocchi and I.~Turcanu, ``Age of information of one-hop broadcast
  communications in a csma network,'' \emph{IEEE Commun. Letters}, vol.~25,
  no.~1, pp. 294--298, 2020.

\bibitem{vinel2015vehicle}
A.~Vinel, L.~Lan, and N.~Lyamin, ``{Vehicle-to-vehicle communication in
  C-ACC/platooning scenarios},'' \emph{IEEE Commun. Mag.}, vol.~53, no.~8, pp.
  192--197, 2015.

\bibitem{AOI_DCC}
N.~{Lyamin}, B.~{Bellalta}, and A.~{Vinel}, ``{Age-of-Information-Aware
  Decentralized Congestion Control in VANETs},'' \emph{IEEE Netw. Letters},
  vol.~2, no.~1, pp. 33--37, 2020.

\bibitem{yates2018age}
R.~D. Yates and S.~K. Kaul, ``{The age of information: Real-time status
  updating by multiple sources},'' \emph{IEEE Trans. on Info. Theory}, vol.~65,
  no.~3, pp. 1807--1827, 2018.

\bibitem{Kosta_2019_AOI_packet_management}
A.~Kosta, N.~Pappas, A.~Ephremides, and V.~Angelakis, ``{Age of information
  performance of multiaccess strategies with packet management},'' \emph{J.
  Commn. Netw.}, vol.~21, no.~3, pp. 244--255, 2019.

\bibitem{Maatouk_2020_AOI}
A.~Maatouk, M.~Assaad, and A.~Ephremides, ``{On the Age of Information in a
  CSMA Environment},'' \emph{IEEE/ACM Trans. Netw.}, vol.~28, no.~2, pp.
  818--831, 2020.

\bibitem{Cao_WCNC_AOI_2022}
L.~{Cao}, H.~{Yin}, R.~{Wei}, and L.~{Zhang}, ``Optimize semi-persistent
  scheduling in nr-v2x: An age-of-information perspective,'' in \emph{2022 IEEE
  Wireless Communications and Networking Conference (WCNC)}, 2022, pp.
  2053--2058.

\bibitem{Challenges_CV2X_Gyawali_2021}
S.~Gyawali, S.~Xu, Y.~Qian, and R.~Q. Hu, ``Challenges and solutions for
  cellular based v2x communications,'' \emph{IEEE Commun. Surv. Tutor.},
  vol.~23, no.~1, pp. 222--255, 2021.

\bibitem{kaul2011minimizing}
S.~Kaul, M.~Gruteser, V.~Rai, and J.~Kenney, ``{Minimizing age of information
  in vehicular networks},'' in \emph{Proc., IEEE Commun. Soc. Conf. Sens. Mesh
  Ad Hoc Commun. Netw.}\hskip 1em plus 0.5em minus 0.4em\relax IEEE, 2011, pp.
  350--358.

\bibitem{LTEV2Vsim_paper}
G.~Cecchini, A.~Bazzi, B.~M. Masini, and A.~Zanella, ``{LTEV2Vsim: An LTE-V2V
  simulator for the investigation of resource allocation for cooperative
  awareness},'' in \emph{Proc., IEEE Int. Conf. Models Technol. Intell. Transp.
  Syst. MT-ITS}, 2017, pp. 80--85.

\end{thebibliography}

\end{document}

% 	\subsection{IoT}

% \section{Section 3}
% \label{section3}
% \input{sections/3_section.tex}

% \section{Section 4}
% \label{section4}
% \input{sections/4_section.tex}
   
% %example for Bullet point list

% \begin{itemize}
% \item example
% \end {itemize}

% %example for numbered list
%     \begin{enumerate}
%     \item example
 
%     \end{enumerate}\textbf{Input:} All slots in sensing history, $sf_{total} (1:N_{\rm sensing})$, Power threshold selected from previous slot selection ($P_{th}^{t-1}$), RRI selected from previous AoI scheduling instance, RRI$^{t-1}$, $cap_{th}$ \\